\pdfoutput=1 
\newcommand{\figurespath}{.}

\RequirePackage{fix-cm}
\documentclass[twocolumn,epjc3]{svjour3} 
\smartqed  

\RequirePackage{graphicx}

\usepackage{subfigure}
\usepackage{mathrsfs}
\usepackage[T1]{fontenc} 
\usepackage{graphicx}
\usepackage{subfigure}
\usepackage{amsmath}

\usepackage{lineno}

\usepackage{xspace}
\usepackage{placeins}
\usepackage{afterpage}

\usepackage{verbatim}
\usepackage{cite}
\usepackage{url}
\usepackage[hyperindex,breaklinks]{hyperref}

\usepackage{preprintcover}  
\PreprintCoverPaperTitle{A measurement of the ratio of the production cross sections for $W$ and $Z$ bosons in association with jets with the ATLAS detector}  
\PreprintIdNumber{CERN-PH-EP-2014-200}  
\PreprintCoverAbstract{The ratio of the production cross sections for 
$W$ and $Z$ bosons in association with jets has been measured in
proton--proton collisions at $\sqrt{s}=7\,\mathrm{TeV}$ with the ATLAS
experiment at the Large Hadron Collider.
The measurement is based on the entire 2011 dataset, 
corresponding to an integrated luminosity of
$4.6\,\mathrm{fb}^{-1}$.  Inclusive and differential cross-section
ratios for massive vector bosons decaying to electrons and muons are
measured in association with jets with transverse momentum
$p_\mathrm{T} >30\,\mathrm{GeV}$ and jet rapidity $|y| < 4.4$.  The
measurements are compared to next-to-leading-order perturbative QCD
calculations and to predictions from different Monte Carlo generators
implementing leading-order matrix elements supplemented by parton
showers.}  
\PreprintJournalName{Eur. Phys. J. C}  

\usepackage{amsmath}
\usepackage{graphicx}
\usepackage{subfigure}

\usepackage{booktabs}

\usepackage{lscape}
\usepackage{rotating}

\usepackage{fp}
\usepackage{xparse} 
\ExplSyntaxOn
\DeclareExpandableDocumentCommand \mult {  m m  }
 {
  \fp_eval:n { round( 100 * #1 / #2, 1 ) }
 }
\ExplSyntaxOff

\newcommand{\rjetsabstract} {
The ratio of the production cross sections for 
$W$ and $Z$ bosons in association with jets has been measured in
proton--proton collisions at $\sqrt{s}=7\,\mathrm{TeV}$ with the ATLAS
experiment at the Large Hadron Collider.
The measurement is based on the entire 2011 dataset, 
corresponding to an integrated luminosity of
$4.6\,\mathrm{fb}^{-1}$.  Inclusive and differential cross-section
ratios for massive vector bosons decaying to electrons and muons are
measured in association with jets with transverse momentum
$p_\mathrm{T} >30\,\mathrm{GeV}$ and jet rapidity $|y| < 4.4$.  The
measurements are compared to next-to-leading-order perturbative QCD
calculations and to predictions from different Monte Carlo generators
implementing leading-order matrix elements supplemented by parton
showers.}

\newcommand{\TeV}{\ifmmode {\mathrm{\ Te\kern -0.1em V}}\xspace\else
                   \textrm{Te\kern -0.1em V}\xspace\fi}%

\newcommand{\GeV}{\ifmmode {\mathrm{\ Ge\kern -0.1em V}}\xspace\else
                   \textrm{Ge\kern -0.1em V}\xspace\fi}%

\newcommand{\met}{\ensuremath{E_{\mathrm{T}}^{\mathrm{miss}}}\xspace} 

\newcommand{\alp}{ALPGEN\xspace}

\newcommand{\her}{HERWIG\xspace}
\newcommand{\pyt}{PYTHIA\xspace}
\newcommand{\she}{SHERPA\xspace}
\newcommand{\bh}{{\sc BlackHat}\xspace}
\newcommand{\bhs}{{\sc BlackHat}+SHERPA\xspace}

\newcommand{\jim}{JIMMY\xspace}

\newcommand{\ace}{{\sc Acer}MC\xspace}
\newcommand{\pow}{POWHEG-BOX\xspace}
\newcommand{\per}{PERUGIA2011C\xspace}
\newcommand{\ctt}{CT10\xspace}
\newcommand{\cts}{CTEQ6L1\xspace}

\newcommand{\gea}{GEANT4\xspace}

\newcommand{\njets}{\ensuremath{{N}_{\mathrm{jets}}}\xspace}

\newcommand{\ifb}{\ensuremath{\rm fb^{-1}}\xspace}
\newcommand{\ipb}{\ensuremath{\rm pb^{-1}}\xspace}
\newcommand{\mt}{\ensuremath{m_{\mathrm{T}}}\xspace}

\newcommand{\dr}{\ensuremath{\Delta R}\xspace}

\newcommand{\tev}{\ifmmode {\mathrm{\ Te\kern -0.1em V}}\else
                   \textrm{Te\kern -0.1em V}\fi}%
\newcommand{\gev}{\ifmmode {\mathrm{\ Ge\kern -0.1em V}}\else
                   \textrm{Ge\kern -0.1em V}\fi}%
\newcommand{\el}{\ensuremath{e}\xspace}

\newcommand{\htj}{\ensuremath{H_{\rm T}}\xspace}
\newcommand{\stj}{\ensuremath{S_{\rm T}}\xspace}

\newcommand{\pt}{\ensuremath{p_{\rm T}}\xspace}

\newcommand{\ptj}{\ensuremath{p^{\rm j}_{\rm T}}\xspace}
\newcommand{\ayj}{\ensuremath{|y|}\xspace}

\newcommand{\mll}{\ensuremath{m_{\ell\ell}}\xspace}

\newcommand{\Zmm}{\ensuremath{Z\to\mu\mu}\xspace}

\newcommand{\Zee}{\ensuremath{Z\to\el\el}\xspace}
\newcommand{\Zll}{\ensuremath{Z\to\ell\ell}\xspace}
\newcommand{\Zjets}{\ensuremath{\Zg\,\texttt{+}\,\mathrm{jets}}\xspace}
\newcommand{\Zmmjets}{\ensuremath{\Zgmm\,\texttt{+}\,\mathrm{jets}}\xspace}
\newcommand{\Zeejets}{\ensuremath{\Zgee\,\texttt{+}\,\mathrm{jets}}\xspace}
\newcommand{\Zlljets}{\ensuremath{\Zgll\,\texttt{+}\,\mathrm{jets}}\xspace}

\newcommand{\Wjets}{\ensuremath{W\,\texttt{+}\,\mathrm{jets}}\xspace}

\newcommand{\antibar}[1]{\ensuremath{#1\bar{#1}}}
\newcommand{\ttbar}{\antibar{t}\xspace}

\newcommand{\Wln}{\ensuremath{W\to\ell\nu}\xspace}
\newcommand{\Wlnjets}{\ensuremath{W(\to\ell\nu)\,\texttt{+}\,\mathrm{jets}}\xspace}
\newcommand{\Wenjets}{\ensuremath{W(\to\el\nu)\,\texttt{+}\,\mathrm{jets}}\xspace}
\newcommand{\Wmnjets}{\ensuremath{W(\to\mu\nu)\,\texttt{+}\,\mathrm{jets}}\xspace}

\newcommand{\Zg}{\ensuremath{Z}\xspace}

\newcommand{\Zgll}{\ensuremath{\Zg\,(\to\ell \ell)}\xspace}
\newcommand{\Zgee}{\ensuremath{\Zg\,(\to\el\el)}\xspace}
\newcommand{\Zgmm}{\ensuremath{\Zg\,(\to\mu\mu)}\xspace}

\newcommand{\antikt}{anti-$k_{t}$\xspace}

\newcommand{\rjets}{\ensuremath{R_{\mathrm{jets}}}\xspace}

\newcommand{\tight}{``tight''\xspace}

\newcommand{\dressed}{``dressed''\xspace}

\journalname{Eur. Phys. J. C} 

\newcommand{\paper}{paper~} 

\begin{document} 

\title{A measurement of the ratio of the production cross sections for $W$ and $Z$ bosons in association with jets with the ATLAS detector}

\titlerunning{Ratio of $W$ and $Z$ boson production cross sections in association with jets} 

\author{The ATLAS Collaboration} 

\institute{CERN, 1211 Geneva 23, Switzerland \label{addr1}} 

\date{Received: date / Accepted: date} 

\maketitle 

\begin{abstract} 
\rjetsabstract 
\end{abstract} 


\section{Introduction \label{sec:Intro}} 

Precise measurements of the production of vector bosons in association with jets
are important tests of quantum chromodynamics (QCD) and provide constraints
on background processes to Higgs boson studies and to searches for new physics.
The measurement of the ratio of \Wjets to \Zjets\footnote{In 
this \paper, $W$ means a $W^{+}$ or $W^{-}$~boson and 
$Z$ is defined as a 
$Z$ 
or $\gamma^{*}$~boson.} production cross sections, termed \rjets, 
directly probes the difference between the kinematic distributions of the jet system 
recoiling against the $W$ or $Z$ bosons.

In comparison to separate \Wjets and \Zjets cross section measurements, the  
\rjets measurement is a more precise test of  perturbative QCD (pQCD), 
since some experimental uncertainties and
effects from non-perturbative processes,
such as hadronization and multi-parton
interactions, are greatly reduced in the ratio.
This allows precise comparisons with state-of-the-art Monte
Carlo simulations and next-to-leading-order (NLO) perturbative QCD calculations to be made.

At low energies, the difference in vector-boson masses
translates to a change in momentum transfer between incoming partons 
and thus different hadronic radiation patterns. 
In addition, the parton distribution
functions of the proton (PDFs) imply different quark--gluon and quark--antiquark
contributions to \Wjets and \Zjets processes. 

At very high energies, the vector-boson mass difference is not large
relative to the momentum transfer, so differences between \Wjets
and \Zjets production are expected to decrease, even though some differences in
the parton distribution functions remain.
A precise measurement of \rjets can therefore be used, in the
context of searches for new particles or interactions beyond the
Standard Model, to infer the \Wjets contribution, given \Zjets production in the same phase space, 
or vice versa.
The \rjets measurement may also be sensitive to direct contributions from new
particle production, if
the new particles decay via $W$ or $Z$ bosons~\cite{Abouzaid:2003is}.
New physics phenomena are generally expected to appear in various topologies with 
high-momentum jets or high jet multiplicities, highlighting the importance of 
studying QCD effects in those regions of phase space.
 
The ATLAS collaboration
performed the first measurement of \rjets as a function of the
jet transverse momentum in events with exactly one jet 
in proton--proton collisions at
$\sqrt{s}=7\,\TeV$, 
using a data sample corresponding to an integrated luminosity of $33\,\ipb$~\cite{Aad:2011xn}.
This result demonstrated that the precision obtained in such a
measurement is sufficient to be sensitive to the QCD effects mentioned
above.
The CMS collaboration
performed an \rjets measurement of the jet multiplicity in vector-boson production with 
up to four associated jets, based on a 
 similar dataset corresponding to an integrated luminosity of 
$36\,\ipb$  in $pp$ collisions collected at
$\sqrt{s}=7\,\TeV$~\cite{Chatrchyan:2011ne}.
The results reported in this \paper are based on a dataset
corresponding to an integrated luminosity of
$4.6\,\ifb$, collected with the ATLAS detector during the 
2011 $pp$ collision run of the LHC at $\sqrt{s} = 7\,\TeV$.
This dataset is over a hundred times larger than the one used
in previously published results, allowing improved precision
over a much larger region of phase space as well as the study of previously
inaccessible differential distributions.

The \rjets measurement is done for the electron
and muon decay channels of the $W$ and $Z$ bosons 
for jets with transverse momentum $\pt > 30\,\GeV$ and rapidity 
$|y| < 4.4$.\footnote{ATLAS uses a right-handed coordinate system with its origin
  at the nominal interaction point (IP) in the centre of the detector
  and the $z$-axis along the beam pipe.  The $x$-axis points from the
  IP to the centre of the LHC ring, and the $y$-axis points upward.
  Cylindrical coordinates ($r$, $\phi$) are used in the transverse
  plane, $\phi$ being the azimuthal angle around the beam pipe.  The
  pseudorapidity is defined in terms of the polar angle $\theta$ as
  $\eta = -\ln \tan(\theta/2)$.}
 The measurements of the electron and muon channels are performed in
slightly different phase spaces and combined in a common phase space
defined in terms of the $\pt$ and pseudorapidity $\eta$ of the leptons, 
the invariant mass of the 
$Z$ boson, the angular separation between the two leptons\footnote{
\label{fn:deltar}
Angular
separations
between particles or reconstructed objects are measured in $\eta$--$\phi$ space using\\
$
\Delta R \equiv
\sqrt{ (\Delta\phi)^{2} \texttt{+} (\Delta \eta)^{2} }
$. 
}
 of the $Z$ boson decay, and the transverse mass\footnote{\label{fn:mt}
The transverse mass of the $W$ boson is reconstructed as
$m_T =\sqrt{2\ensuremath{p_{\mathrm{T}}}^{\ell}\ensuremath{p_{\mathrm{T}}}^{\nu}(1-\cos(\phi^{\ell}-\phi^{\nu}))}$
where $p_{\mathrm{T}}^{\ell}$ and $p_{\mathrm{T}}^{\nu}$ are the transverse momenta of the charged lepton and
the neutrino respectively and $\phi^{\ell}$ and $\phi^{\nu}$ their azimuthal directions. 
}
of the $W$ boson, as presented in Table~\ref{tab:comb}.
\begin{table*} 
\renewcommand{\arraystretch}{1.3}
\centering
\begin{tabular}{l|l}
\toprule
 Lepton $\pt$ and pseudorapidity $\eta$ & $\pt > 25\,\GeV$, $|\eta| < 2.5$ \\
 $W$ transverse mass and neutrino $\pt$  & $\mt > 40\,\GeV$, $\pt > 25\,\GeV$ \\
 $Z$ invariant mass and 
 lepton--lepton angular separation & $66 < \mll < 116\,\GeV$, $\dr_{\ell \ell}>0.2$ \\
Jet $\pt$, 
  rapidity and jet--lepton angular separation  & $\pt > 30 \GeV$, $|y| < 4.4$, $\dr_{j\ell} > 0.5$ \\
\bottomrule
\end{tabular}
\caption{
Particle-level phase space of the present \rjets measurement.
\label{tab:comb}
}
\centering
\end{table*}
The $W$ and $Z$ selections are based on the  \Wjets and \Zjets cross-section measurements detailed in Ref.~\cite{ATLASWjets2014,Aad:2013ysa}, 
with 
a minor  update in the $Z$ selection to further reduce the uncertainty on the \rjets measurement.
In the results reported here, 
\rjets is measured as a function of the inclusive and exclusive
jet multiplicity (\njets) up to four jets. An extensive set of
differential measurements is also presented, in which \rjets is measured as a
function of the transverse momentum and the rapidity of the leading jet, which is the 
one with largest transverse momentum, in 
events with at least one jet.
The ratio \rjets is also presented as a function of the 
transverse momentum and rapidity of the second and third leading jets in 
events with at least two or three jets respectively.
A set of differential measurements as a function of dijet observables
in events with at least two jets is presented. The measurement
of \rjets as a function of the summed scalar \pt of the jets (\stj) for
different jet multiplicities is also reported.  The results are compared
to several Monte Carlo generators and with next-to-leading-order 
pQCD predictions corrected for non-perturbative effects.

The \paper is organized as follows. 
The experimental setup is described in Sect.\ \ref{sec:Detector}.  
Section\ \ref{sec:MonteCarlo} provides
details on the simulations used in the measurement, and
Sect.\ \ref{sec:Selection} discusses the event selection. 
The estimation of background contributions is described in
Sect.\ \ref{sec:Background}, and 
the procedure used to correct the measurements for detector 
effects is described in Sect.\ \ref{sec:Unfolding}. 
The treatment of the systematic uncertainties is described in Sect.\ \ref{sec:Systematics}. 
Section\ \ref{sec:Combination} discusses the combination of the
electron and muon results.
Section\ \ref{sec:Theory} provides details
on the NLO pQCD predictions.
Finally, Sect.\ \ref{sec:Results} discusses the results, and 
Sect.\ \ref{Conclusions} presents the conclusions.

\section{The ATLAS detector \label{sec:Detector}}

The ATLAS detector~\cite{Aad:2008zzm} is a multi-purpose detector with a
symmetric cylindrical geometry and nearly $4\pi$
coverage in solid angle. 
The
collision point is surrounded by inner tracking devices followed by a
superconducting solenoid providing a 2~T magnetic field, a calorimeter
system, and a muon spectrometer.  The inner tracker provides precision
tracking of charged particles for pseudorapidities $|\eta| < 2.5$.  It
consists of 
silicon pixel and microstrip detectors and a straw-tube transition radiation tracker.
The calorimeter system has liquid argon (LAr) or
scintillator tiles as active media.  In the pseudorapidity region
$|\eta| < 3.2$, high-granularity LAr electromagnetic (EM) sampling
calorimeters are used.  An iron/scintillator tile calorimeter provides
hadronic coverage for $|\eta| < 1.7$.  The endcap and forward regions,
spanning $1.5<|\eta| <4.9$, are instrumented with LAr calorimeters for
both the EM and hadronic measurements.  The muon spectrometer consists of
three large superconducting toroids, 
each comprising eight coils,
and a system
of trigger chambers and precision tracking chambers that provide
triggering and tracking capabilities in the ranges $|\eta| < 2.4$ and
$|\eta| < 2.7$ respectively.

The ATLAS trigger system uses three consecutive levels. 
The Level-1 triggers are hardware-based and use coarse detector information to 
identify regions of interest,
whereas the Level-2 triggers are based on fast online data reconstruction algorithms.
Finally, the Event Filter triggers use offline data reconstruction algorithms. 

\section{Monte Carlo simulation \label{sec:MonteCarlo}}

Simulated event samples were used to correct the measured distributions for
detector effects and acceptance, to determine some background
contributions and to correct theory calculations for non-perturbative effects.
Signal samples of $W (\rightarrow \ell \nu, \ell= e, \mu) \texttt{+}$jets and $Z
(\rightarrow \ell \ell, \ell= e, \mu) \texttt{+}$jets events were generated with
\alp~\\ v2.13~\cite{Mangano:2002ea},
with up to five additional partons in the final state.
It was interfaced to \her~v6.520~\cite{Corcella:2000bw} for
parton showering and fragmentation, with \jim~v4.31
~\cite{Butterworth:1996zw} for contributions from multi-parton interactions
and with PHOTOS~\cite{Golonka:2006tw} to calculate final-state QED
radiation.  
The \mbox{\cts~\cite{Pumplin:2002vw}} PDFs were used with
the \mbox{AUET2-}\mbox{CTEQ6L1} tune~\cite{ATLAS:2011gmi}, a set of
specific non-perturbative event generation parameter values.
Similar samples were produced with \alp~v2.14 interfaced
to \pyt v6.425~\cite{Sjostrand:2006za} using the \per~\cite{perugia}
tune and PHOTOS. They were used to estimate the uncertainties on non-perturbative
corrections for parton-level NLO pQCD predictions.
An additional set of signal samples was
generated with {\she}~v1.4.1~\cite{Gleisberg:2008ta,Hoeche:2012yf} and \ctt
PDFs~\cite{Lai:2010vv}. 
Top quark pair production (\ttbar) was simulated with ALPGEN and HERWIG+JIMMY, 
in the same configuration as for the signal samples. 
Additional \ttbar samples were generated with the \pow 
generator v1.0~\cite{Alioli:2010xd}, 
using the {CT10} next-to-leading order (NLO) PDFs
and interfaced to \pyt~v6.425.
These additional samples were reserved for the
evaluation of the systematic uncertainties.
Single top-quark production, including $Wt$ production, 
was modelled with {\ace}~3.8~\cite{Kersevan:2004yg}
interfaced to \pyt 
and MRST LO* PDFs~\cite{Sherstnev:2007nd}.
The diboson production processes $W^+W^-, WZ$,
and $ZZ$ were generated with \her~v6.510 and \jim~v4.3
using the MRST LO* PDFs~\cite{Sherstnev:2007nd} and
the {\sc AUET2-LO*} tune~\cite{ATLAS:2011gmi}.

The generated Monte Carlo (MC) samples
were overlaid with additional inelastic $pp$ scattering events
generated with \pyt v6.425, following the distribution of the average
number of $pp$ interactions in the selected data.
The samples were then passed through the simulation of the ATLAS
detector based on \gea~\cite{Agostinelli:2002hh, Aad:2010ah} and
through the related trigger simulation.  

All samples were normalized to the inclusive cross section calculated
at the highest pQCD order available.
The $W/Z\texttt{+}$jets signal samples were normalized to
the next-to-next-to-leading-order (NNLO) pQCD inclusive Drell--Yan
predictions calculated with the FEWZ~\cite{Melnikov:2006kv} program and the
MSTW2008 NNLO PDFs~\cite{Martin:2009iq}.
The \ttbar samples were normalized to the cross section calculated at NNLO+NNLL in
Refs.~\cite{Cacciari:2011hy, Baernreuther:2012ws, Czakon:2012zr, Czakon:2012pz, Czakon:2013goa,Czakon:2011xx}, 
and the diboson samples were normalized to cross
sections calculated at NLO using
{\sc MCFM}~\cite{Campbell:2011bn} with the MSTW2008
PDF set.

The simulated events were reconstructed and analysed with the same
analysis chain as the data.
Scale factors were applied to the simulated samples to correct the 
lepton trigger, reconstruction, and identification efficiencies to match those measured in data.

\section{Event selection \label{sec:Selection}}
\begin{table*}
\renewcommand{\arraystretch}{1.3}
\centering
\begin{tabular}{lll} 
\toprule 
  & Electron selection & Muon selection \\ 
\midrule 
Lepton \pt & $\pt > 25$~\GeV & $\pt > 25$~\GeV \\ 
Lepton pseudorapidity & $|\eta|<2.47$ (excluding $1.37<|\eta|<1.52$) &$|\eta|<2.4$ \\
\midrule
  \multicolumn{3}{c}{\Wln\ event selection} \\ 
\midrule $Z$ veto & \multicolumn{2}{l}{exactly one selected lepton} \\ 
Missing transverse momentum  & \multicolumn{2}{l}{$\met > 25$~\GeV} \\ 
Transverse mass & \multicolumn{2}{l}{$\mt > 40$~\GeV} \\ 
\midrule
\multicolumn{3}{c}{\Zll\ event selection} \\
\midrule 
Multiplicity & \multicolumn{2}{l}{exactly two selected leptons} \\ 
Charge & \multicolumn{2}{l}{opposite sign} \\ 
Invariant mass & \multicolumn{2}{l}{$66 < m_{\ell\ell} < 116$~\GeV} \\ 
Separation & \multicolumn{2}{l}{$\Delta R_{\ell\ell}>0.2$} \\ 
\midrule
  \multicolumn{3}{c}{Jet selection} \\ 
\midrule 
Transverse momentum & \multicolumn{2}{l}{$\pt > 30$~\GeV} \\ 
Jet rapidity & \multicolumn{2}{l}{|$y| < 4.4$} \\ 
Jet--lepton angular separation & \multicolumn{2}{l}{$\dr_{\ell {\rm j}} > 0.5$} \\
\bottomrule
\end{tabular}
\caption{Kinematic event selection criteria for 
 \Wlnjets and \Zlljets
 event samples.
\label{tab:selection}}
\centering
\end{table*}
The data samples considered in this \paper correspond to a total integrated
luminosity of $4.6\,\ifb$, with an uncertainty of 1.8$\%$~\cite{Aad:2013ucp}.
Table~\ref{tab:selection} summarizes the kinematic requirements for
leptons, $W$ bosons, $Z$ bosons, and jets.  
The selection criteria for
$W$ boson candidates were defined using the largest possible coverage
of the ATLAS detector for electrons, muons and jets.  
The selection criteria for $Z$ boson candidates were modified with respect
to those in Ref.~\cite{Aad:2013ysa}, to be as similar as possible to the $W$
 boson selection in order to maximize the cancellation of uncertainties in the \rjets 
measurement: triggers requiring at least one lepton were employed,
the minimum lepton transverse momentum was raised from
20~\GeV to 25~\GeV, tighter criteria were used to identify electrons
and slightly looser requirements were placed on the second leading lepton
with respect to the leading one.

The data were collected using single-electron or single-muon 
triggers, employing the same requirements for the $W$ and $Z$ data selections. 
Electron-channel events
were selected using a trigger that required the presence of at least
one electron candidate, formed by an energy cluster consistent with an
electromagnetic shower in the calorimeter and associated to an inner
detector track.  
Electron candidates were required to have a
reconstructed transverse energy above $20\,\GeV$ or $22\,\GeV$, depending on
the trigger configuration of the different data periods.
Muon-channel events were recorded using a trigger
that required the presence of at least one muon candidate
with transverse momentum above $18\,\GeV$.  
Lepton trigger
thresholds were low enough to ensure that leptons with $\pt > 25\,\GeV$
lie on the trigger efficiency plateau.

Events were required to have a primary vertex, defined as the vertex in
the event with the highest summed $\pt^2$ of
all associated tracks, among vertices with at least three tracks.

Electrons were reconstructed by matching clusters of energy found in
the electromagnetic calorimeter to tracks reconstructed in the inner
detector. Candidate electrons had to satisfy the 
\tight quality requirements 
defined in Ref.~\cite{Aad:2014fxa},
which include requirements on the calorimeter shower shape, track
quality, and association of the track with the energy cluster found in the calorimeter.
Electron candidates had to  have $\pt > 25\,\GeV$ and $|\eta |<2.47$, where the
transition region between barrel and endcap electromagnetic
calorimeter sections at $1.37<|\eta|<1.52$ was excluded.

Muons were reconstructed from track segments in the muon spectrometer
that were matched with  tracks in the inner
detector~\cite{Aad:2014zya}, and were required to  have $\pt > 25$~\GeV
and $|\eta|<2.4$.  
To suppress particles from hadron decays, the leading  muon had to be consistent with originating from the primary vertex by requiring
$|d_0/\sigma(d_0)| < 3.0$, where $d_0$ is the transverse impact parameter of the muon and
$\sigma(d_0)$ is its uncertainty.

In order to suppress background from multi-jet events where a jet is misidentified as a lepton,
the leading
lepton was required to be isolated.  An additional \pt- and $\eta$-dependent
requirement on a combination of calorimeter and track isolation variables 
was applied to the leading electron, in order to yield a constant efficiency across different momentum ranges and detector regions, as detailed in Ref.~\cite{ele_iso}.
The track-based isolation uses a cone size
 of $\Delta R = 0.4$ 
and the calorimeter-based isolation uses a cone size of $\dr = 0.2$.
The actual isolation requirements range between 2.5~\GeV and 4.5~\GeV for the calorimeter-based isolation and between 2.0~\GeV and 3.0~\GeV for the track-based isolation.
For muon candidates, the
scalar sum of the transverse momenta of tracks within a cone
of size $\Delta R = 0.2$ around the leading muon 
had to  be less than $10\%$ of its
transverse momentum.

Reconstructed $W$ candidates were required to have exactly one
selected lepton. The missing transverse momentum in the event had to
have a magnitude \met greater than $25$~\GeV, and the transverse mass
$m_T$ had to be greater than $40$~\GeV.  The magnitude and azimuthal
direction of the missing transverse momentum are measured from the
vector sum of the transverse momenta of calibrated physics objects and
additional soft calorimeter deposits~\cite{Aad:2012re}.  Reconstructed
$Z$ candidates were required to have exactly two selected leptons of
the same flavour with opposite charge.  Their invariant mass
$m_{\ell\ell}$ had to be in the range 
$66 \leq m_{\ell\ell} \leq 116\,\GeV$ 
and the leptons had to be separated by $\Delta
R_{\ell\ell} > 0.2$.

Jets were reconstructed using the 
\antikt
algorithm~\\ \cite{Cacciari:2008gp} with a distance parameter $R=0.4$ on
topological clusters of energy in the calorimeters~\cite{Aad:2012ag}.  
Jets were required to have a
transverse momentum above $30 \gev$ and a rapidity of $\ayj < 4.4$.
Jets within $\dr =0.5$ of a selected lepton were removed.  
The energy and the direction of reconstructed jets were corrected 
to account for the point of origin, assumed to be the primary vertex, and
for the bias introduced by the presence of additional $pp$ interactions in the same bunch crossing (``pile-up''). 
 The jet energy was then calibrated to account for the different response of the calorimeters to electrons and hadrons and for energy losses in un-instrumented regions by applying correction factors derived from 
simulations. 
A final calibration, derived from in-situ techniques using Z+jet balance, 
$\gamma$+jet balance and multi-jet balance, was applied to the data 
to reduce residual differences between data and simulations~\cite{Aad:2013tea}.

In order to reject jets from  pile-up,
a jet selection was applied based on
the ratio of the summed scalar \pt\ of tracks originating from the
primary vertex and associated with the jet to the summed \pt of all tracks associated with the jet.
Jets were selected if this ratio was above $0.75$.
This criterion was applied to jets within $|\eta|<2.4$, so that
they are inside the inner tracker acceptance.
Comparison between data and 
simulation for various data periods
confirmed that the residual impact of pile-up 
on the distribution of the jet observables
in this analysis is well modelled by the simulation.  

The numbers of \Wjets and \Zjets candidate events in the electron and muon channels 
for each jet multiplicity are shown in Tables~\ref{tab:electronZ} and \ref{tab:muonZ}, 
together with the corresponding numbers of predicted events. The expected fraction of predicted events from signal and each background source, determined as described in the next section, is also shown.

\section{Background estimation \label{sec:Background}}

Background processes to $W$ and $Z$ boson production associated with jets
can be classified into three categories.
The first category, referred to as electroweak background, consists of
diboson production, 
vector-boson production with subsequent decay to $\tau$-leptons, 
and
``cross-talk'' background, in which the signal 
\Wjets (\Zjets) production appears as background in the \Zjets (\Wjets) sample.
These background contributions are relatively small
(about 10\% in the \Wjets electron channel, 
about 6\% in the \Wjets muon channel, and about 1\% in \Zjets, 
as shown in Tables 
\ref{tab:electronZ} and
\ref{tab:muonZ})
and were thus estimated using simulated event samples.

The second category consists of events where the leptons are produced
in decays of top quarks. 
The \ttbar component
completely dominates the background contribution to \Wjets events at
high jet multiplicities, amounting to approximately $20\%$ of the
sample with 
\ensuremath{W\texttt{+}\geq\mathrm{3\, jets}}
and increasing to approximately $45\%$ 
for events with four selected jets.
The effect is less dramatic in \Zjets events, where the \ttbar
background contributes 
about $5\%$ to the sample of events with
\ensuremath{Z\texttt{+}\geq\mathrm{3\, jets}}
and about $10\%$ to the 
sample with four jets.
The background contribution from single top-quark production
is 
about $4\%$ of the sample in \Wjets events 
for events with three or four jets, and smaller at lower jet multiplicities.
This contribution is even smaller in \Zjets events.
Contributions from \ttbar events to \Wjets candidates with at least
three jets, where this background dominates, were estimated with a
data-driven method as described below in order to reduce the overall
uncertainty.
The \ttbar contributions to \Wjets candidates with fewer than three 
jets and to \Zjets events were estimated using simulated event
samples, as are the contributions from single top quarks.

The third category of background, referred to as multi-jet background, comes from events 
in which hadrons mimic the signature of an isolated lepton.
In the electron channel this includes photon conversion processes, typically
from the decay of neutral pions, narrow hadronic jets and real
electrons from the decay of heavy-flavour hadrons. In the muon
channel, the multi-jet background is primarily composed of heavy-flavour hadron decay processes. This background category dominates at
low jet multiplicity in \Wjets events, amounting to $11 \%$ of the selected 
sample
in both the electron and muon channels
for events with one jet. Data-driven techniques were 
used to estimate this background contribution to both the \Wjets and
\Zjets candidate events, as described below.
The methods employed to estimate background contributions with data-driven
techniques in this analysis are very similar between candidate events with $W$ bosons and $Z$ bosons
and between electron and muon channels.

\paragraph{\ttbar background} \hspace{0pt}

\

The \ttbar background is the dominant background contribution to
\Wjets events with at least three jets, since each top quark predominantly decays
as $t \rightarrow Wb$.  
The size of the \ttbar contribution was estimated with a maximum-likelihood fit to the data.

The \ttbar template in this fit was derived from 
  a top--quark-enhanced data sample
by requiring, in addition to the selection criteria given in Table~\ref{tab:selection}, 
at least one $b$-tagged jet in the event,
as determined by the MV1 $b$-tagging algorithm of Ref.~\cite{btag}.
The chosen MV1 algorithm working point has a $b$-tagging efficiency of 70$\%$.
This data sample is contaminated with $W$ signal events and electroweak and multi-jet backgrounds,
amounting to about 40$\%$ in events with three jets and 25$\%$ in events with four jets.
The contribution from $W$ signal events and electroweak background
was estimated using 
simulation. The multi-jet contribution
to the top-enriched sample was estimated using the multi-jet background estimation method as outlined 
in the last part of this section, 
but with an additional $b$-tagging requirement.
Potential biases in the 
\ttbar templates extracted from data
were investigated using 
simulated \ttbar events. 
Since $b$-tagging is only available 
for jets within $|\eta| < 2.4$ where information from the tracking detectors exists, 
the $b$-tagging selection biases some of the kinematic distributions, most notably the jet rapidity distribution. To account for this, \alp \ttbar~simulations 
were used to correct for any residual bias in the differential distributions; the maximum correction is 30\%.

The number of \ttbar events 
was extracted by fitting a discriminant distribution to the sum of three
templates: the top-enriched template after subtracting the
contaminations discussed above, the multi-jet template (determined as
described below) and the template obtained from simulation of
the \Wjets signal and the other background sources.  The chosen
discriminant was the transformed aplanarity, given by $\exp(-8A)$,
where ${A}$ is the aplanarity defined as 1.5 times the smallest
eigenvalue of the normalized momentum tensor of the leptons and all
the jets passing the selection~\cite{Aad:2012qf}.
This discriminant provides the best separation 
between \ttbar and the \Wjets signal.
The fit to the transformed aplanarity distribution was done in the range $0.0$--$0.85$
in each exclusive jet multiplicity of three or more.

Since the top-enriched sample is a sub-sample of the signal sample,
statistical correlation between the two samples is expected.  Its size
was estimated using pseudo-datasets by performing Poisson variations
of the signal and top-enriched samples.  To account for this
correlation, the uncertainty on the fit was increased by 15$\%$ for
events with three jets and about 30$\%$ for events with four jets.

\paragraph{Multi-jet background} \hspace{0pt}

\

The multi-jet background contribution to the \Wjets selected events was 
estimated with a template fit method using 
a sample enriched in multi-jet events.
The templates of the multi-jet background for  the fit were extracted from data, by  
modifying the lepton isolation requirements in both the electron and 
muon channels, in order  to select  non-isolated leptons. The
templates of the signal, the \ttbar background,
and the electroweak background were obtained from 
simulation.
These templates
were then normalized by a fit to the $\met$ distribution 
after all signal requirements other than the requirement on \met 
were applied. 

To select an electron-channel data sample enriched in multi-jet events, 
dedicated electron triggers based on loose requirements were used 
(as defined in Ref.~\cite{Aad:2014fxa}),
along with additional triggers based on loose electron and jet selection criteria.
The background 
template distributions were built from events for which the 
identification requirements of the nominal electron selection 
failed, in order to suppress signal contamination in the template. 
Candidate electrons were also required to be
non-isolated in the calorimeter, i.e.\ were required to have an energy
deposition in the calorimeter in a cone of size $\dr < 0.3$ centred on their direction 
greater than 20\% of 
their total transverse energy.
This selection
results in a data sample highly enriched in
jets misidentified as electrons.
As the luminosity increased during the course of 2011, the trigger selections 
were adjusted to cope with the increasing 
trigger rates. 
In order to build multi-jet template distributions that provide a good representation of the
pile-up conditions of the selected data sample, these template distributions were
extracted from two distinct
data periods with high and low pile-up conditions. 
The background templates extracted from the two different data periods were fitted 
separately and then combined into an overall multi-jet estimate.

To select the multi-jet sample in the muon channel,
 muon candidates were required to be non-isolated.
The sum of transverse momenta of  tracks in a cone of size $\dr < 0.2$ centred on
the muon-candidate direction had to be between $10\%$ and $50\%$ of the muon transverse momentum. 
The contamination from  $W$ signal events and electroweak and top backgrounds to the 
multi-jet sample was subtracted using 
simulation. 
It amounts to $1.4 \%$ for events with one jet and $4.8 \%$ for events with  four jets.

The number of multi-jet background events was obtained for each jet multiplicity
in the electron and muon channels
by fitting the \met distribution obtained from the \Wjets data
candidate events (selected before the application of the \met
requirement) to the multi-jet template and a template of signal and
electroweak and \ttbar backgrounds derived from simulations.
The fit range was chosen to ensure significant contributions from both templates, 
in order to guarantee fit stability under systematic variations described 
in Sect.~\ref{sec:Systematics}. 
The \met distribution was fitted in the range $15\, \GeV$ to $80\,\GeV$ in the electron channel 
and in the range $15\, \GeV$  to $70\,\GeV$ in the muon channel.  

The multi-jet background contribution to the \Zjets selected candidates was 
estimated using 
a template fit method 
similar
to the procedure used in the  \Wjets case.
In the
electron channel, the template distributions for the multi-jet background
were  constructed from a data sample collected
with  electron triggers looser than those used for the nominal \Zee selection.
Electrons were then required to satisfy  the loose offline identification criteria 
(as defined in Ref.~\cite{Aad:2014fxa})
but fail to meet the 
nominal
criteria.
In the muon channel, the multi-jet  template distributions for the multi-jet background were 
obtained from the nominal signal data sample, after relaxing the
impact parameter significance requirement applied to \Zmm events
candidates, and selecting events that did not satisfy  the isolation criteria 
applied in the signal selection.
The number of multi-jet background events was obtained for each exclusive jet multiplicity
by fitting the dilepton invariant mass distribution
 $m_{\ell\ell}$ in an extended range,
 $50 < m_{\ell\ell} < 140$~\GeV, 
 excluding the $Z$-peak region itself, after all other signal requirements were applied. 
Due to statistical limitations for jet multiplicities greater than two
jets, the normalisation factor obtained from the two-jet bin was 
consistently applied to the templates for higher jet multiplicities.
Potential bias in this procedure was accounted for in the systematic uncertainty estimate.

The evaluation of the systematic uncertainties for each background source is explained in Sect.~\ref{sec:Systematics}. 

\begin{table*}
\footnotesize

  \renewcommand{\arraystretch}{1.1}
    \centering
    \begin{tabular}{%
l|ccccc
    }
    \toprule
$N_{\mbox{jets}} $ & \mbox{0}  & \mbox{1} & \mbox{2}  &  \mbox{3}  & \mbox{4}  \\ 
    \midrule    
Fraction [\%] & \multicolumn{5}{c}{ $W (\rightarrow e\nu)$ + jets } \\ 
    \midrule    
$W\rightarrow e\nu$ & 94  &  78	&  73	& 58   & 37 \\
$Z\rightarrow ee$ & 0.30	& 7.5	& 6.6	& 6.8	& 5.4	  \\
$t\bar{t}$  		& \multicolumn{1}{r}{\mbox{$<0.1\;\ $}}  
					& 0.30 & 3.4		& 18	& 46 \\
Multi-jet   		& 4 	& 11 	& 12		&  11	& 6.9 \\
Electroweak (without $Z\rightarrow ee$)	& 1.9 	& 2.6 		& 3.3	& 3	& 1.9 \\
Single top  		& \multicolumn{1}{r}{\mbox{$<0.1\;\ $}} 
				    & 0.30 		& 	1.7	& 3.5     & 3.9 \\
    \midrule    
 Total Predicted  & 11 100 000& 1 510 000 & 354 000& 89 500 &  28 200\\
                  & $\pm$ 640 000 & $\pm$ 99 000 & $\pm$ 23 000 & $\pm$ 5600   & $\pm$ 1400  \\  
 Data Observed    & 10 878 398  & 1 548 000 & 361 957  &  91 212 &  28 076   \\

    \midrule
    
Fraction [\%] &  \multicolumn{5}{c}{ $Z (\rightarrow ee) $ + jets } \\
    \midrule        
$Z\rightarrow ee$ & 100  & 99	&  96 & 93	& 90    \\
$W\rightarrow e\nu$ & \multicolumn{1}{r}{\mbox{$<0.1\;\ $}} & \multicolumn{1}{r}{\mbox{$<0.1\;\ $}} & \multicolumn{1}{r}{\mbox{$<0.1\;\ $}} 
					& \multicolumn{1}{r}{\mbox{$<0.1\;\ $}} & \multicolumn{1}{r}{\mbox{$<0.1\;\ $}} \\
$t\bar{t}$ 			& \multicolumn{1}{r}{\mbox{$<0.1\;\ $}}  		
				    & 0.20 	& 1.9 	&  4.6	  & 7.8 \\
Multi-jet  			&  	0.20	& 0.20 	&  0.40		& 0.50	  & 0.50 \\
Electroweak (without $W\rightarrow e\nu$) & 0.10 		&  0.50	& 1.3 	& 1.4  & 1.2 \\
Single top                     & \multicolumn{1}{r}{\mbox{$<0.1\;\ $}} & \multicolumn{1}{r}{\mbox{$<0.1\;\ $}} &   0.10  & 0.20 & 0.10 \\
   \midrule
 Total Predicted 	&  754 000 & 96 500 & 22 100 & 4700 & 1010 \\
                 	&  $\pm$	47 000 & $\pm$ 6900  & $\pm$ 1700 & $\pm$ 930  & $\pm$ 93 \\
 Data Observed 		& 761 280 & 99 991 & 22 471 & 4 729 & 1050 \\
   \bottomrule
\end{tabular}
\caption{
The contribution of signal and background from various sources, expressed as a fraction of the total number of expected events for the 
\Wenjets and \Zeejets selection as a function of jet multiplicity \njets together with the total numbers of expected and observed events.   
}
\label{tab:electronZ}
 
\end{table*}

\vspace{1cm}

\begin{table*}
\footnotesize

  \renewcommand{\arraystretch}{1.1}
    \centering
    \begin{tabular}{%
l|ccccc
    }
    \toprule
$N_{\mbox{jets}} $ & \mbox{0}  & \mbox{1} &  \mbox{2}  & \mbox{3}  & \mbox{4} \\
    \midrule    
 Fraction [\%] &  \multicolumn{5}{c}{ $W (\rightarrow \mu\nu)$ + jets } \\ 
    \midrule
$W\rightarrow \mu\nu$ & 93 	&  82	& 78  & 62 & 40 \\  
$Z\rightarrow \mu\mu$ & 3.4 & 	3.5  	& 3.5 & 3 & 2 \\
$t\bar{t}$ 			  & \multicolumn{1}{r}{\mbox{$<0.1\;\ $}} %
					  & 0.20		& 3.1 & 19 & 46 \\%
Multi-jet  			  & 1.5 		& 	11	& 10 & 9.5 & 6.8 \\ 
Electroweak (without $Z\rightarrow \mu\mu$) & 	1.9	& 	2.7	&3.4 & 2.9 & 1.9 \\%
Single top 			  & \multicolumn{1}{r}{\mbox{$<0.1\;\ $}} %
					  &  0.20		&  1.7 & 3.4 & 3.8 \\%
\midrule 
 Total Predicted 	  & 13 300 000& 1 710 000 	& 384 000 & 96 700 	& 30 100 \\%
		  	  & $\pm$ 770 000	    & $\pm$ 100 000	&  $\pm$	24 000  &  $\pm$ 6100       & $\pm$  1600\\%
 Data Observed 		  & 13 414 400 & 1 758 239 	& 403 146 & 99 749  & 30 400 \\
\midrule
 Fraction [\%] &  \multicolumn{5}{c}{ $Z (\rightarrow \mu\mu)$ + jets } \\
\midrule 

$Z\rightarrow \mu\mu$ 	&  100 & 	99&  96 &  91 & 84 	\\%
$W\rightarrow \mu\nu$ 	& \multicolumn{1}{r}{\mbox{$<0.1\;\ $}} %
				        & 0.10 		&  0.10	&  0.20	&  0.20 \\%
$t\bar{t}$ 			  	& \multicolumn{1}{r}{\mbox{$<0.1\;\ $}} 
				        &  0.30	&  2.2	&  6.1	&  13	\\%
Multi-jet  				&  0.30		& 0.50	&  0.90	& 	1.1 & 1.7\\%
Electroweak (without $W\rightarrow \mu\nu$)&  0.10		& 0.50 		& 1.3 	&1.4  & 1.1 	\\
Single top 				& \multicolumn{1}{r}{\mbox{$<0.1\;\ $}} %
						& \multicolumn{1}{r}{\mbox{$<0.1\;\ $}} %
					& 0.10	&  0.20	&  0.20 	\\
\midrule 
 Total Predicted 		& 1 300 000	& 168 000 & 37 800 	&  8100	& 1750 \\
 				&$\pm$ 79 000	& $\pm$ 12 000	& $\pm$ 2800	& $\pm$ 660 & $\pm$ 160  \\%
 Data Observed 			& 1 302 010 	& 171 200 & 38 618 	& 8397 	& 1864 \\%
\bottomrule 
\end{tabular}
\caption{ 
The contribution of signal and background from various sources, expressed as a fraction of the total number of expected events for the 
\Wmnjets and \Zmmjets selection as a function of jet multiplicity \njets together with the total numbers of expected and observed events.   
}
\label{tab:muonZ}

\end{table*}

\section{Corrections for detector effects \label{sec:Unfolding}}

The signal event yields were determined by
subtracting the estimated background contributions from the data.
After background subtraction, the resulting distributions were 
corrected for detector effects such that distributions at
particle level were obtained.  
The correction procedure based on simulated samples corrects for
jet, $W$ and $Z$ selection efficiency, resolution effects and residual mis-calibrations.
While
\Wjets and \Zjets events were separately corrected before forming \rjets,
the systematic uncertainties were estimated for 
the ratio itself, as explained in the next section.

At particle level, the lepton kinematic variables in the MC-generated
samples were computed using final-state leptons from the $W$ or $Z$
boson decay.  Photons radiated by the boson decay products within a
cone of size \dr\ $= 0.1$ around the direction of a final-state lepton
were added to the lepton, and the sum is referred to as the \dressed
lepton.  Particle-level jets were identified by applying the \antikt
algorithm with $R$ = 0.4 to all final-state particles with a lifetime
longer than $30\,\rm{ps}$, whether produced directly in the
proton--proton collision or from the decay of particles with shorter
lifetimes.  Neutrinos, electrons, and muons from decays of the $W$ and
$Z$ bosons, as well as collinear photons included in the ``lepton
dressing procedure'' were excluded by the jet reconstruction
algorithm.  The phase-space requirements match the selection criteria
defining the data candidate events, as presented in
Table~\ref{tab:selection}, in order to limit the dependence of the
measurement results on theoretical assumptions.

The correction was implemented using an iterative Bayesian method of
unfolding~\cite{D'Agostini:1994zf}.
Simulated events are 
used to generate for each distribution a response matrix to account for bin-to-bin migration effects
 between the reconstruction-level and particle-level distributions.
The Monte Carlo particle-level prediction is used as initial prior 
to determine a first estimate of the unfolded data distribution. 
For each further iteration, the previous estimate of the
unfolded distribution is used as a new input prior.
Bin sizes in each distribution were 
chosen to be a few times larger than the resolution of the
corresponding variable. 
The \alp \Wjets and \Zjets samples provide a
satisfactory description of distributions in data and were employed to
perform the correction procedure.  
The number of iterations was optimized to find a balance between too many iterations, 
causing high statistical uncertainties associated with the unfolded spectra, 
and too few iterations, which increase the dependency on the Monte Carlo prior. 
The optimal number of iterations is typically between one and three, 
depending on the observable. Since the differences in the unfolded results 
are negligible over this range of iterations, 
two iterations were used  consistently for unfolding each observable.

\section{Systematic uncertainties \label{sec:Systematics}}

\begin{table*}
\centering
\begin{tabular}{l|lllll}
\toprule
\multicolumn{ 6 }{c}{  $ (W \rightarrow e\nu) / (Z \rightarrow ee)$ } \\
        $N_\mathrm{jets}$ & $\ge 0$  & $\ge 1$  & $\ge 2$  & $\ge 3$  & $\ge 4$  \\
\midrule
{ Electron} & 0.89 & 0.92 & 0.93 & 0.97 & 1.0 \\
{ JES} & 0.094 & 2.0 & 2.0 & 3.5 & 5.7 \\
{ JER} & 0.25 & 2.4 & 3.5 & 4.3 & 6.4 \\
{ \met} & 0.19 & 1.7 & 1.2 & 1.2 & 1.0 \\
{ $t\bar{t}$} & 0.024 & 0.23 & 1.0 & 4.9 & 14 \\
{ Multi-jet} & 0.81 & 1.6 & 1.5 & 2.2 & 6.2 \\
{ Other backgrounds} & 0.12 & 0.57 & 0.58 & 0.76 & 1.0 \\
{ Unfolding } & 0.20 & 0.56 & 0.86 & 1.2 & 1.4 \\
{ Luminosity } & 0.062 & 0.26 & 0.27 & 0.34 & 0.44 \\
\midrule
{ Total } & 1.3 & 4.1 & 4.8 & 8.2 & 18 \\
\bottomrule
\end{tabular}
\vspace{0.5cm}
\begin{tabular}{l|lllll}
\toprule
\multicolumn{ 6 }{c}{  $ (W \rightarrow \mu\nu) / (Z \rightarrow \mu\mu)$ } \\
         $N_\mathrm{jets}$ & $\ge 0$  & $\ge 1$  & $\ge 2$  & $\ge 3$  & $\ge 4$  \\
\midrule
 { Muon} & 1.1 & 1.2 & 1.1 & 0.86 & 0.87 \\
 { JES} & 0.10 & 0.84 & 0.71 & 1.8 & 2.6 \\
 { JER} & 0.094 & 1.6 & 1.8 & 2.6 & 4.2 \\
 { \met} & 0.30 & 1.0 & 0.94 & 0.97 & 0.99 \\
 { $t\bar{t}$} & 0.018 & 0.18 & 0.87 & 4.3 & 12 \\
 { Multi-jet} & 0.20 & 0.60 & 1.1 & 1.7 & 2.7 \\
 { Other backgrounds} & 0.21 & 0.24 & 0.28 & 0.42 & 0.60 \\
 { Unfolding } & 0.22 & 0.59 & 0.90 & 1.2 & 1.2 \\
 { Luminosity } & 0.10 & 0.12 & 0.11 & 0.088 & 0.023 \\
 \midrule
 { Total } & 1.2 & 2.5 & 3.0 & 5.9 & 13 \\
 \bottomrule
 \end{tabular}
 \caption{Systematic uncertainties  in percent on the measured \Wjets\ / \Zjets\
 cross-section ratio in the electron and muon channels as a function of
 the inclusive jet multiplicity \njets.}
\label{tab:systematics}
\end{table*}

One of the advantages of measuring \rjets is that systematic
uncertainties that are positively correlated between the numerator and
denominator cancel at the level of their correlations (higher
correlations result in larger cancellations).  The impact on the ratio
of a given source of uncertainty was estimated by simultaneously
applying the systematic variation due to this source to both the \Wjets
and \Zjets events and repeating the full measurement chain with the
systematic variations applied.  This included re-estimating the
data-driven background distributions after the variations had been
applied.

Since the uncertainties  were found to be symmetric within the statistical fluctuations, the resulting systematic uncertainties on the \rjets measurements were fully symmetrized by taking 
the average value of 
the upwards and downwards variations.

Uncertainty sources affecting the \rjets measurements can be assigned
to one of the following categories: jet measurements, lepton
measurements, missing transverse momentum measurement, unfolding
procedure, data-driven background estimates and simulation-based
background estimates.  These sources of uncertainty feature
significant correlations between \Wjets and \Zjets processes, which
have been fully accounted for as explained above.  The systematic
uncertainties on the \ttbar and multi-jet background estimates were
considered to be uncorrelated between the \Wjets and \Zjets
selections.  The uncertainty on the integrated luminosity was
propagated through all of the background calculations and treated as
correlated between \Wjets and \Zjets so that it largely cancels in the
ratio.  The contributions from each of the sources mentioned above and
the total systematic uncertainties were obtained by adding in
quadrature the different components, and are summarized in
Table~\ref{tab:systematics}.  The total uncertainty on \rjets as a
function of the inclusive jet multiplicity ranges from $4\%$ for $\njets\geq
1$ to 18\% for $\njets\geq 4$ in the electron channel and from 3\% for
$\njets \geq 1$ to 13\% for $\njets\geq 4$ in the muon channel.

Jet-related systematic uncertainties are dominated by the uncertainty on the jet energy 
scale (JES) and resolution (JER).
The JES uncertainty was derived via in-situ calibration techniques, 
such as the transverse momentum
balance in \Zjets,  multi-jet and $\gamma-$jet events, for which
a comparison between data and 
simulation was performed~\cite{Aad:2013tea}.
The JER uncertainty  was derived from
a comparison of the resolution measured in dijet data events using
the bisector method~\cite{Aad:2012ag}, and the same approach was applied to
simulated
dijet events.
The JER and JES uncertainties are highly correlated between \Wjets and \Zjets
observables and are thus largely suppressed
compared to the individual measurements. They are
nevertheless the dominant systematic uncertainties in the
cases where there are one or two jets in the events.
The  cancellation  is not perfect because 
any changes in JES and JER are consistently propagated to the 
\met measurement event-by-event. This causes larger
associated migrations for the $W$ selection  than for the $Z$ selection.  
In addition, the level of background in the \Wjets sample is larger, 
resulting in a larger jet uncertainty compared to the \Zjets selection.
The sum of JER and JES uncertainties on the \rjets measurement ranges from 3\% to 8\% in the electron channel
and from 2\% to 5\% in the muon channel as $\njets$ ranges from 1 to 4.
The difference between the two channels is due to the fact that the $Z\to ee$ background
in the $W \to e\nu$ data sample is much larger than the 
corresponding $Z\to \mu\mu$ background
in the $W \to \mu \nu$ sample, being about 
7\% for candidate events with one jet in the electron channel compared to 3\% in the muon channel. 
The $Z\to ee$ background contaminates the $W \to e\nu$ sample
because one electron can be misidentified as a jet, contributing to the JES and JER uncertainties.
This contribution to the uncertainties does not cancel in \rjets.

The uncertainty on the electron and muon selections includes 
uncertainties on the electron energy or muon momentum scale and 
resolution, as well as uncertainties on the scale factors applied to the 
simulations in order to 
match the electron or muon trigger, reconstruction and identification efficiencies to those in data. 
Any changes in
lepton energy scale and resolution were consistently
propagated to the \met   measurement.
The energy- or momentum-scale corrections of the leptons were 
obtained from comparison of the $Z$-boson invariant mass distribution 
between data and simulations. 
The uncertainties on the scale factors have 
been derived from a comparison of tag-and-probe results in data
and simulations~\cite{Aad:2014fxa,Aad:2014zya}. 
Each of these sources of uncertainty is relatively small in the \rjets measurement 
(about $1\%$ for $N_{\mathrm{jets}}$ ranging from 1 to 4 in both channels).

The uncertainties in \met due to uncertainties in JES, JER, lepton energy scale and resolution 
were included in the values quoted above. A residual \met uncertainty accounts for uncertainties 
on the energy measurement of 
clusters in the calorimeters 
that are not associated with electrons or jets. 
It was determined via in-situ measurements and comparisons between data and 
simulation~\cite{ATLAS:Met}. These systematic uncertainties affect only the 
numerator of  the ratio because no \met cut was applied to $Z \texttt{+}$ jets  
candidate events.  
The resulting uncertainty on the \rjets measurement is 
about $1\%$ for $N_{\mathrm{jets}}$ ranging from 1 to 4 in both channels. 

The uncertainty on the unfolding has a component of statistical origin
that comes from the limited number of events in each bin of the Monte
Carlo inputs.  This component was estimated from the root mean square
of
\rjets measurements
obtained in a large set of pseudo-data generated independently from
the \Wjets and \Zjets Monte Carlo samples used to unfold the data.
The Monte Carlo modelling uncertainty in the unfolding procedure 
was estimated using an alternative set of \alp samples
for which the nominal \Wjets and \Zjets production was modelled by
different theoretical parameter values.  The MLM matching
procedure~\cite{Alwall:2007fs}, employed to remove the double counting
of partons generated by the matrix element calculation and partons
produced in the parton shower, uses a matching cone of size $R=0.4$ for
matrix element partons of $\pt > 20$~\GeV.  To determine how the
choice of this cone size and the matching \pt scale impact the
unfolded results, samples with variations of these parameters were
used in the unfolding procedure.  In addition, to account for the
impact of changing the amount of radiation emitted from hard partons,
\alp Monte Carlo samples were generated with the renormalisation and 
factorisation scales set to half or twice their nominal value of 
$\sqrt{ m_{V}^2\texttt{+}{\pt}_{V}^2}$, where $V$ is
the $W$ or  $Z$ boson depending on the sample.
The systematic uncertainty is the sum in quadrature of the differences
with respect to the \rjets  
measurement obtained from the nominal samples.
The overall uncertainty on the unfolding procedure
ranges between $0.6 \%$ and $1.4 \%$ for $N_{\mathrm{jets}}$ ranging from 1 to 4.

For backgrounds estimated using simulation, the uncertainty on the 
cross-section calculation 
was taken into account.
The combined impact of these uncertainties on the \rjets
measurement  is typically less than 1$\%$ for the different jet multiplicities.

For \ttbar predictions taken from the \alp sample, 
the uncertainty on the cross-section calculation is considered, as well as 
a shape uncertainty 
by comparing to the \pow \ttbar sample. 
The largest contribution to the total
uncertainty from the data-driven \ttbar estimate is from the statistical
uncertainty on the fit.
The systematic uncertainty on the data-driven \ttbar estimate 
also covers uncertainties on the 
contamination of the background template by signal events, on the
choice of  fit range and 
other small uncertainties.
The latter include the uncertainties on the $b$-tagging efficiencies and uncertainties on the bias in the 
\ttbar\ distributions when applying the $b$-tagging.
The uncertainty on the contribution from  
$W\texttt{+}$ heavy-flavour events to the \ttbar  template, modelled by \alp Monte Carlo samples, was evaluated by varying the $W\texttt{+}c$ cross section and the combined $W\texttt{+}cc$ and $W\texttt{+}bb$ cross sections.
The size of the variations is a factor of 0.9 and 1.3 respectively. These factors  were obtained from  fits to the data in two control regions, defined as one or two jets and at least one $b$-tagged jet.
This uncertainty, which is $3 \%$ of the number of \ttbar\ events for $N_{\mathrm{jets}}\geq 3$, is largest at lower jet multiplicities where the contribution from $W\texttt{+}$heavy flavour is most significant.
The upper limit of the fit range in transformed aplanarity was varied from the nominal values of $0.85$ to $0.83$ or $0.87$.
The \ttbar uncertainty dominates
for 
final states with high jet multiplicity
due to its increasing contribution, which does not cancel in \rjets.
It amounts to an uncertainty of $14 \%$ on the \rjets measurement in the electron channel and to 
an uncertainty of $12 \%$ in the muon channel for 
events with at least four jets.

In the evaluation of the multi-jet background systematic uncertainties, 
various sources were taken into account.
For the $W\texttt{+}$jets selection,
the uncertainty on the shape of the template distributions of the multi-jet background was studied by varying 
the lepton isolation requirement and identification definition.  
The  nominal template fit range for \met 
was also varied, within\\ $10$~\GeV up and down from the nominal limits.  
The distributions of the signal template were alternatively modelled by \she instead of \alp
and the difference was taken as an uncertainty. 
The statistical uncertainty on the template
normalisation factor from the fit was also included. 
Finally, to evaluate
the uncertainty on the estimate of the multi-jet background to the 
\Zjets events, the fit ranges 
and the 
modelling of the signal and of the electroweak contamination 
were varied 
in the same way as for the \Wjets events.
The combined impact of these uncertainties on the \rjets 
measurement varies between 2\% and 6\% in the electron
channel and between 1\% and 3\% in the muon channel for $N_{\mathrm{jets}}$ 
ranging from 1 to 4.

\section{Combination of electron and muon channels\label{sec:Combination}}

In order to increase the precision of the \Wjets to \Zjets
differential cross-section ratio measurements 
the results obtained for each
observable in the electron and the muon channels were statistically combined,
accounting for correlations between the sources
of systematic uncertainties affecting each channel.  
Since the electron and muon measurements were performed in different fiducial
regions, bin-by-bin correction factors, estimated using \alp Monte
Carlo samples, were applied to each measured distribution to
extrapolate the measurements to the common phase space 
defined in Table~\ref{tab:comb}.
The corrections to the \rjets measurement are of
the order of $6 \%$  in the electron channel and $1 \%$ in the muon channel.
The uncertainties on the acceptance corrections are 
below $0.5 \%$,
as determined by using \she instead of \alp.  
By comparing distributions computed at LO and NLO, it was checked with 
{\sc MCFM}
that  NLO effects on the extrapolation to the 
common phase space are negligible.
Before the combination was performed, the individual results of the two 
channels were compared to each other after extrapolation; 
the results  are 
compatible within their respective 
uncertainties.

The method of combination used is an averaging procedure described
in Refs.~\cite{Glazov:2005rn,Aaron:2009bp}.  
The  distributions for each observable were  combined separately 
by minimising a $\chi^2$ function which takes into account
the results in the extrapolated electron and muon channels
and all systematic uncertainties on both channels.
The uncertainties on the modelling in the unfolding procedure, the integrated luminosity, the background contributions estimated from simulations except for \Zjets 
and \Wjets backgrounds and  systematic uncertainties on the data-driven  \ttbar estimation were treated as correlated among bins and between channels.
The lepton systematic uncertainties were assumed to be correlated between bins of a given distribution, but uncorrelated between the two lepton channel
measurements. The statistical uncertainties of the data, the statistical uncertainty of the unfolding procedure, and the statistical uncertainty of the \ttbar fit
were treated as uncorrelated among bins and channels.
The systematic uncertainties on multi-jet backgrounds, which contain correlated and uncorrelated components, are also treated as uncorrelated among bins and channels. 
This choice has little impact on the final combined results and was chosen as it is 
slightly more conservative in terms of the total uncertainty of the combined results.
Finally, the uncertainties from the jet energy scale, the jet energy resolution, the \met calculation  and the \Zjets and \Wjets background contributions 
were treated as fully correlated between all bins and do not enter into the combination procedure
to avoid numerical
instabilities due to the statistical component in these
uncertainties.
 For the combined results, each of these uncertainties was taken as 
the weighted average of the corresponding uncertainty on the electron and muon measurements, where the weights are the inverse of the sum in quadrature of all the uncorrelated 
uncertainties that entered in the combination. 
\section{Theoretical predictions \label{sec:Theory}}

The measured distributions of all the observables considered in the
analysis are compared at particle level to various pQCD predictions 
in the phase space defined in Table~\ref{tab:comb}.

Next-to-leading-order pQCD predictions were calculated 
with \bhs~\cite{Berger:2009ep,Berger:2010vm,Berger:2010zx} at parton level
for various parton multiplicities, from zero to four.
In this calculation \bh is used for the computation of the virtual one-loop matrix elements, while
\she is used for the real emission part and the phase-space integration.
The fixed-order calculation is performed at parton level only, without radiation and hadronization effects.
Renormalisation and factorisation scales were 
evaluated at $H_{\rm T}/2$, where \htj is defined as the scalar sum of the
transverse momenta of all stable particles 
in each event.
The PDF set used was CT10~\cite{Lai:2010vv}.  Partons were clustered into
jets using the \antikt algorithm with $R=0.4$.

The effect of uncertainties on the prediction has been computed 
for \rjets, accounting for correlation between the individual \Wjets and
\Zjets  processes. The uncertainties on the theoretical
predictions are significantly reduced in this procedure, with the
statistical uncertainty on the samples often dominating.

Uncertainties on the renormalisation and the factorisation scales were
evaluated by varying these scales independently to half and twice
their nominal value.  The PDF uncertainties were computed from the 
CT10 eigenvectors, derived with the Hessian method at
$68\%$ confidence level~\cite{Lai:2010vv}.  
The changes in \rjets due to these PDF variations were combined and used as the uncertainty.
In
addition, the nominal value of the strong coupling constant,
 $\alpha_{s} = 0.118$, was varied by $\pm 0.0012$, and the impact of
this variation was taken into account in the PDF uncertainty.  All the
uncertainty components mentioned above were then added in quadrature.
The total systematic uncertainty on the prediction ranges from 0.3\%
to 1.8\% for inclusive jet multiplicities ranging from one to four,
and from 2\% to 6\% for leading-jet \pt ranging from 
$30\, \GeV$ to
$700\,\GeV$.

In order to compare the \bhs parton-level predictions with the
measurements at particle 
level, a set of
non-perturbative corrections was applied to the predictions.
Corrections for the underlying event (UE) were calculated using samples
generated with
ALPGEN+HERWIG+JIMMY. The ratio of samples where the UE has
been switched on and off was evaluated in each bin of each
distribution.  Corrections for 
the hadronization of partons to jets were 
computed using similar samples by forming the ratio of distributions
obtained using jets clustered from hadrons versus jets clustered from
partons. 
In \rjets, the hadronization and UE corrections have opposite signs
and are quite small (typically 2\% to 3\% for the exclusive jet
multiplicity), so the overall correction factor is close to unity.
Additional \alp+\pyt samples were used to estimate the
uncertainties due to these non-perturbative corrections, which are typically well below $1 \%$.

 Finally, corrections for QED final-state radiation were 
calculated 
as the ratio of \rjets derived from \dressed leptons to \rjets defined before any final-state photon radiation,
using \alp samples interfaced to PHOTOS. 
These corrections range between 1\% and 2\% 
 for both the electron and the muon channel. 
Systematic uncertainties
were derived by comparing with corrections obtained using \she,
which calculates final-state QED radiation using the 
YFS method~\cite{Yennie:1961ad}.
The differences between the predictions are typically well below 1\%.

Tree-level multi-leg matrix element calculations\\ matched to parton
showering algorithms were obtained from the \alp and \she generators.
These calculations use different PDF sets, matching procedures,
parton shower evolution, and hadronization and multi-parton interaction
modelling, as detailed in Sect.~\ref{sec:MonteCarlo}.
Only statistical uncertainties were considered for these predictions,
which are compared with the BlackHat+Sherpa calculations and the data in 
Sect.~\ref{sec:Results}.

\section{Results and discussion \label{sec:Results}}
The theoretical predictions described in Sect.~\ref{sec:Theory} are
compared to the experimental data unfolded to particle level, as
defined in
Sect.~\ref{sec:Unfolding}.  Individual ratios of
the \bhs, \alp, and \she predictions to unfolded data make it
possible to disentangle the important features of each 
theoretical prediction.
The \rjets results highlight the ability of these Monte Carlo programs
to model the differences between \Zjets and \Wjets processes.

Figure~\ref{fig:njet_incl_r} shows \rjets as a function of
exclusive and inclusive jet multiplicity.
The values are detailed in Tables~\ref{tab:njet_excl_r}
 and \ref{tab:njet_incl_r}, respectively.\footnote{Tabulated values of the results are also available in the Durham HepData Project: \url{http://hepdata.cedar.ac.uk}.}~
The theoretical predictions describe the data fairly well, given the experimental uncertainties, with few exceptions.
At high jet multiplicities, where the effects of hard QCD radiation are tested, the SHERPA prediction is about $1.5$ standard deviations ($1.5 \sigma$) of the
experimental error greater than the measurement.
\bhs is able to
describe  \rjets measured as a function of  exclusive jet multiplicity, within the theoretical uncertainties,
although it is about $1\sigma$ greater than the measurement at high inclusive jet 
multiplicities; this is
expected since it does not include all contributions for events with at least four jets.  
\begin{figure*} 
  \centering
  \includegraphics[width=.46\textwidth]{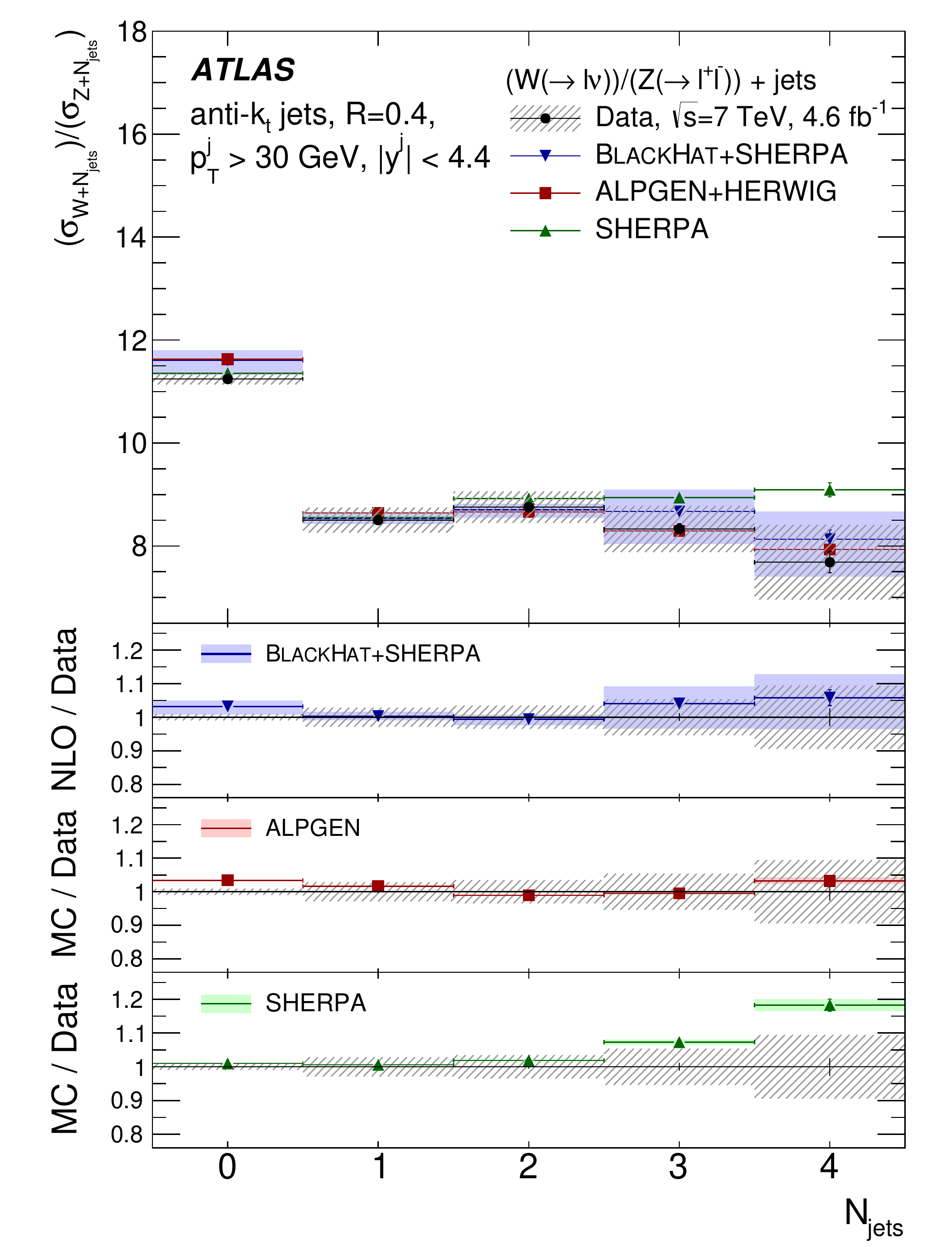}
  \includegraphics[width=.46\textwidth]{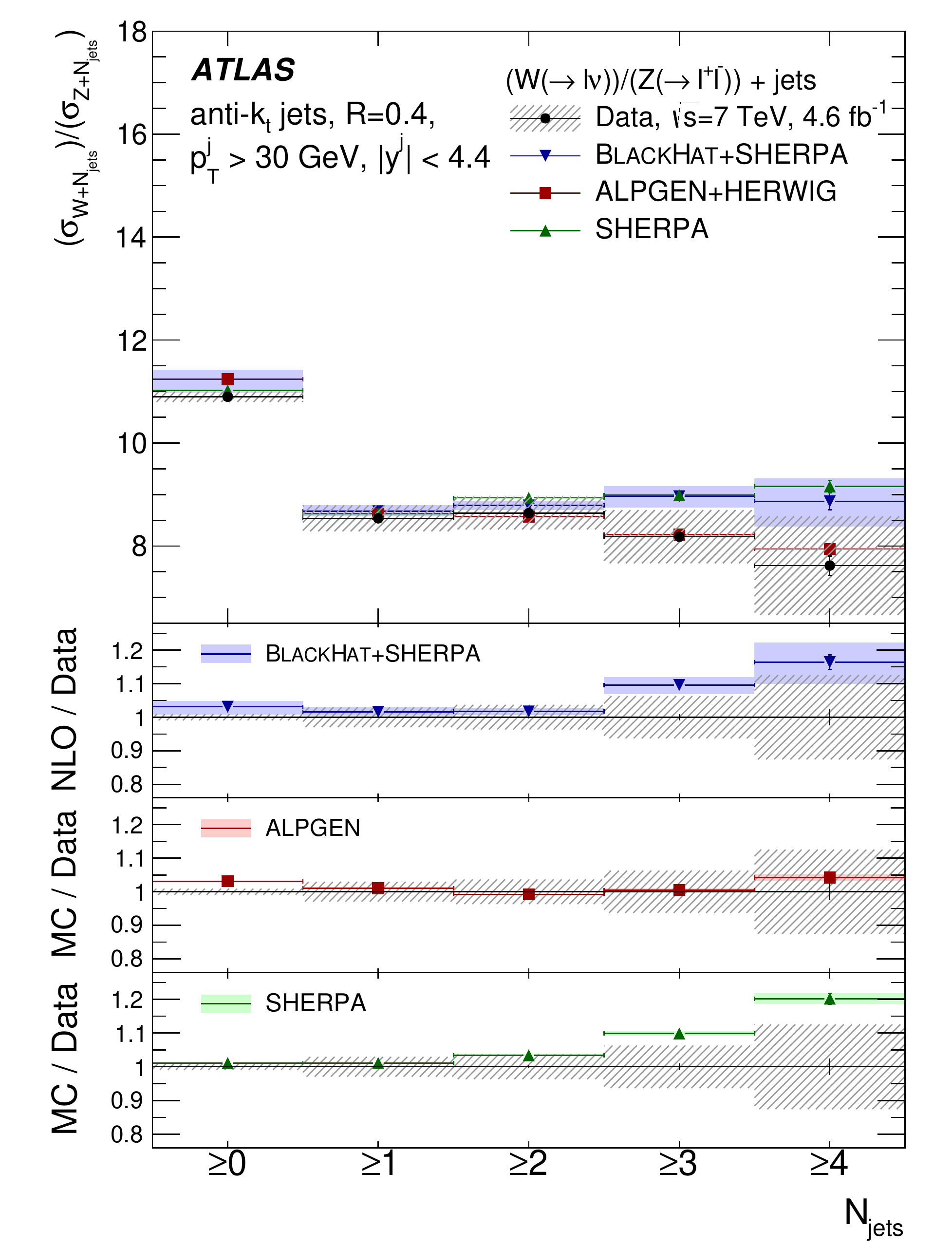}
  \caption{The ratio of \Wjets and \Zjets production cross sections, \rjets, as a function of exclusive jet multiplicity, \njets, 
    (left) and inclusive jet multiplicity (right). The electron and
    muon channel measurements are combined as described in
    the text.  Ratios of the \bhs
    NLO calculation and the {\alp} and {\she} generators to the data are shown in
    the lower panels. 
    Vertical error bars show the respective statistical uncertainties. 
    The hatched error band shows statistical and systematic uncertainties added in quadrature for the data.
    The solid error bands show the statistical uncertainties for the {\alp} and {\she} predictions, and the combined statistical and theoretical uncertainties for the \bhs prediction. 
    }
  \label{fig:njet_incl_r}
\end{figure*}
\begin{table*}  
  \centering 
  \begin{tabular}{%
c|c
  }
  \toprule
  $N_\mathrm{jets}$ &   { \rjets } \\
  \midrule
 = 0  & 11.24  $\,\pm\,$ 0.01 ~(stat.)~   $\,\pm\,$ 0.11 ~(syst.)~ \\
 = 1  & 8.50  $\,\pm\,$ 0.02 ~(stat.)~  $\,\pm\,$ 0.24 ~(syst.)~\\
 = 2  & 8.76  $\,\pm\,$ 0.05 ~(stat.)~  $\,\pm\,$ 0.30 ~(syst.)~\\
 = 3  & 8.33  $\,\pm\,$ 0.10 ~(stat.)~  $\,\pm\,$ 0.44 ~(syst.)~\\
 = 4  & 7.69  $\,\pm\,$ 0.21 ~(stat.)~  $\,\pm\,$ 0.70 ~(syst.)~\\
\bottomrule
\end{tabular}

\caption{The ratio of \Wjets and \Zjets production cross sections, \rjets, 
as a function of exclusive jet multiplicity in the phase space defined in Table~\ref{tab:comb}.}
  \label{tab:njet_excl_r}
\end{table*}
\begin{table*}
  \centering 
  \begin{tabular}{
c|c
  }
  \toprule
  $N_\mathrm{jets}$ &  { \rjets } \\
\midrule
$\geq0$ & 10.90  $\,\pm\,$ 0.01~(stat.)~  $\,\pm\,$ 0.10 ~(syst.)~ \\
$\geq1$ & 8.54  $\,\pm\,$ 0.02~(stat.)~  $\,\pm\,$ 0.25 ~(syst.)~\\
$\geq2$ & 8.64  $\,\pm\,$ 0.04~(stat.)~  $\,\pm\,$ 0.32 ~(syst.)~\\
$\geq3$ & 8.18  $\,\pm\,$ 0.08~(stat.)~  $\,\pm\,$ 0.51 ~(syst.)~\\
$\geq4$ & 7.62  $\,\pm\,$ 0.19~(stat.)~  $\,\pm\,$ 0.94 ~(syst.)~\\
\bottomrule
\end{tabular}
\caption{The ratio of \Wjets and \Zjets production cross sections, \rjets,
as a function of inclusive jet multiplicity in the phase space defined in Table~\ref{tab:comb}.}
  \label{tab:njet_incl_r}
\end{table*}

In the following figures, \rjets is normalized to the ratio
of the $W$ and $Z$  cross sections in the corresponding jet
multiplicity bin presented in Fig.~\ref{fig:njet_incl_r}, so that the 
shapes of the distributions can be compared.
Figure~\ref{fig:jet_pt_r1jet} shows the \rjets
ratio versus the leading-jet \pt for $\njets=1$ and $\njets \ge 1$.
At low transverse momentum ($\pt<200\,\GeV$),
the \rjets distribution falls as the leading-jet \pt increases,
indicating that the shapes in \Wjets and \Zjets events are different.
This is due to the $W$ and $Z$ boson mass difference, which affects the scale of the
parton radiation, 
and the different vector-boson polarizations,
which affect the kinematics of their decay products.  
In the small region very close to the minimum value of the jet \pt considered in the analysis,
where radiative parton shower effects play a major role, 
all of the predicted shapes exhibit trends different from those in the data,
but the \alp predictions still show the best agreement.

\begin{figure*} \centering
  \includegraphics[width=.38\textwidth]{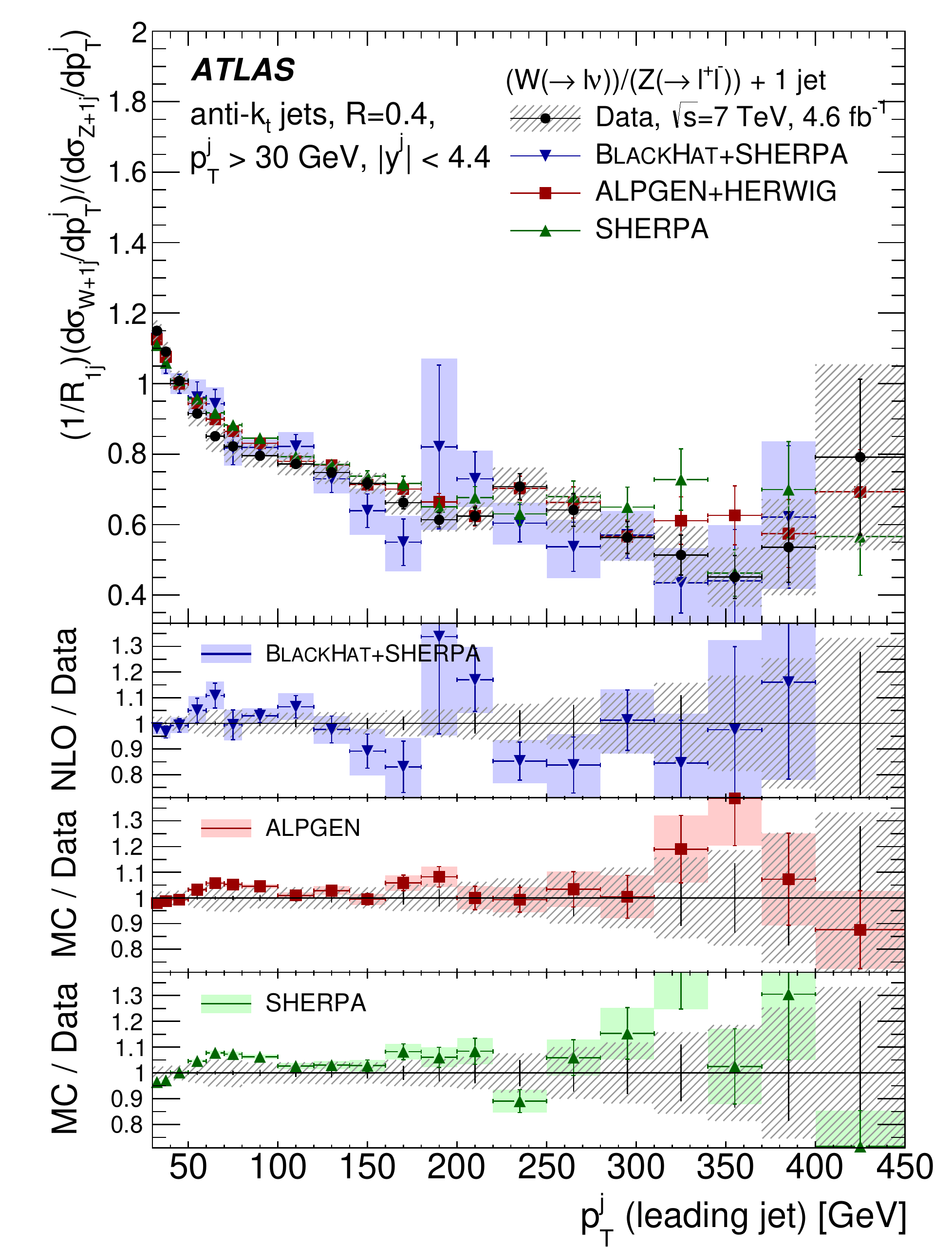}
  \includegraphics[width=.38\textwidth]{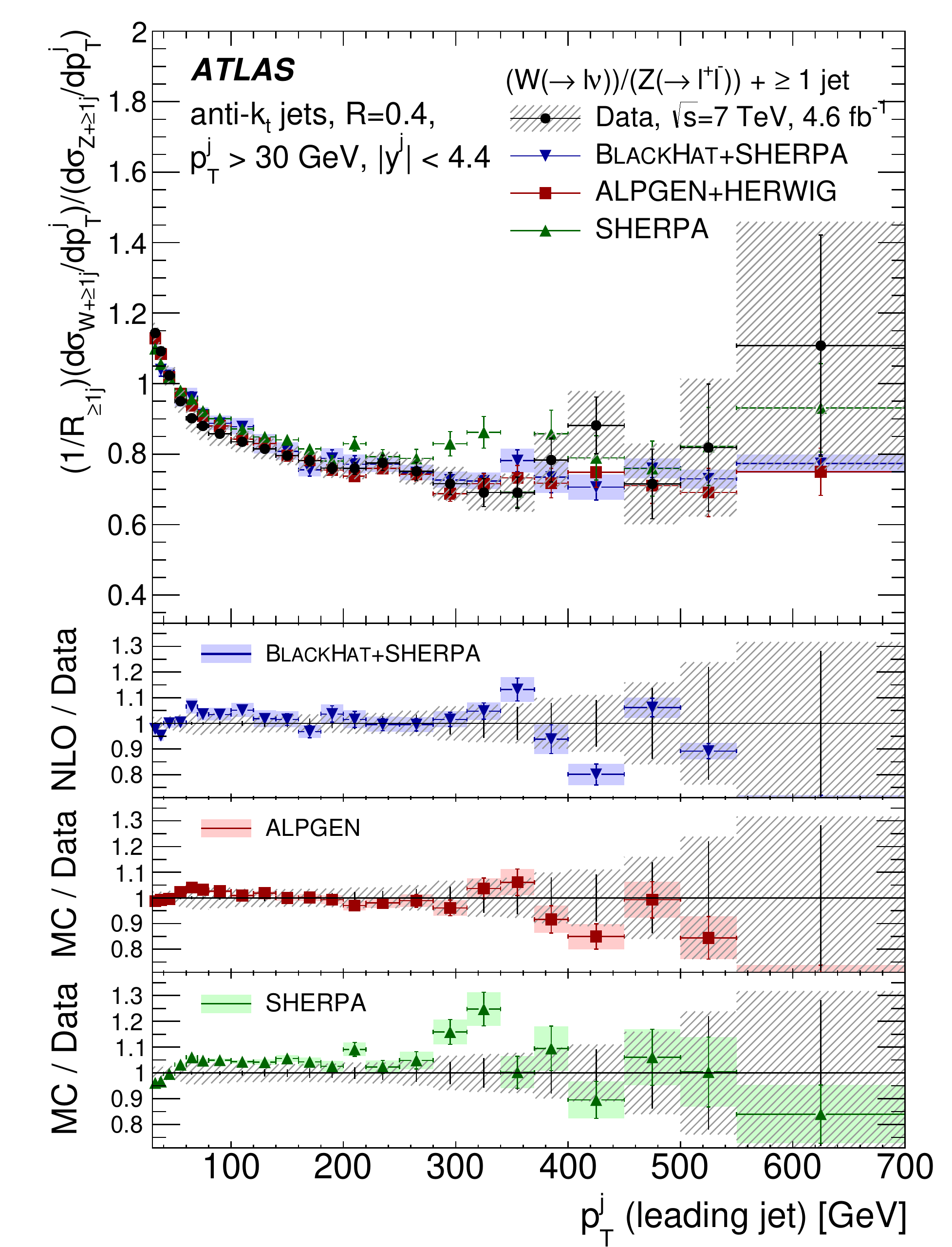}
  \caption{The ratio of \Wjets and \Zjets production cross sections, \rjets, 
  normalized as described in the text versus the leading-jet transverse momentum, \ptj, 
for \njets= 1 (left) and
    \njets$\ge$ 1 (right). The electron and muon channel measurements
    are combined as described in  the text.
    Ratios of the \bhs
    NLO calculation and the {\alp} and {\she} generators to the data are shown in
    the lower panels.
    Vertical error bars show the respective statistical uncertainties. 
    The hatched error band shows statistical and systematic uncertainties added in quadrature for the data.
    The solid error bands show the statistical uncertainties for the {\alp} and {\she} predictions, and the combined statistical and theoretical uncertainties for the \bhs prediction. 
    }
  \label{fig:jet_pt_r1jet}
\end{figure*}

Figure~\ref{fig:jet_pt_r23jets} shows \rjets versus
the leading-jet \pt for \njets$\ge 2$ and \njets$\ge 3$.
The \rjets distribution falls less steeply the more jets are in the event.
This is due to the smaller average vector-boson \pt,
which reduces the effects arising from differences in boson masses and
polarizations.  
At the lowest \pt values considered the comparison with the data shows a tendency for 
different behaviour of the theoretical predictions, 
especially in events with at least three jets. 
The effect, which is most pronounced for \bhs, is expected in case of 
lack of resummation of soft and collinear parton emissions, as in this calculation.

\begin{figure*} 
 \centering
 \includegraphics[width=.38\textwidth]{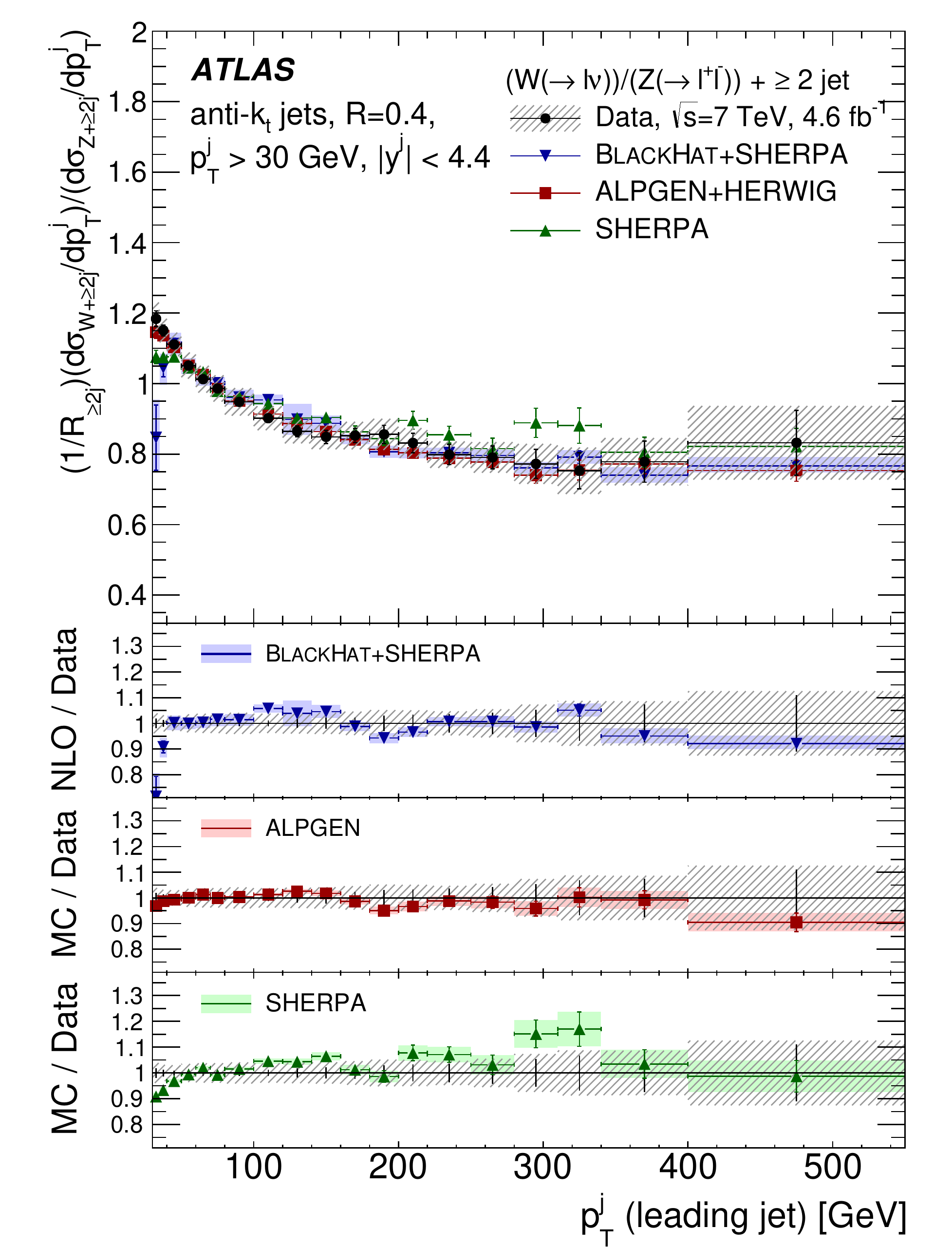}
 \includegraphics[width=.38\textwidth]{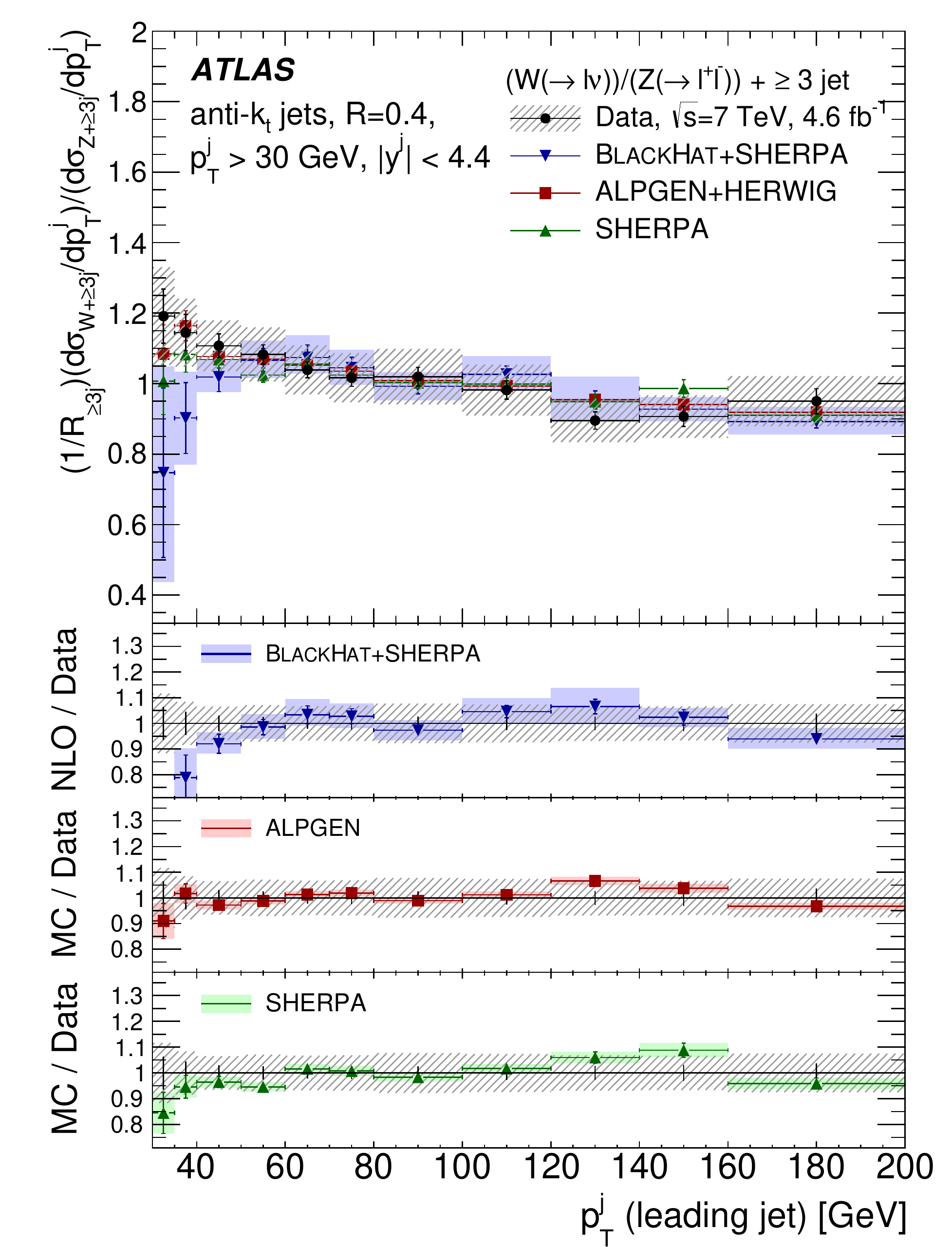}
 \caption{The ratio of \Wjets and \Zjets production cross sections, \rjets, 
 normalized as described in the text versus the leading-jet transverse momentum, \ptj, for \njets$\ge 2$ (left) and
   $\ge 3$ (right). The electron and muon channel measurements are
   combined as described in the text. Ratios
   of the \bhs NLO calculation and the {\alp} and {\she} generators to the data 
   are shown in the lower panels.
    Vertical error bars show the respective statistical uncertainties. 
    The hatched error band shows statistical and systematic uncertainties added in quadrature for the data.
    The solid error bands show the statistical uncertainties for the {\alp} and {\she} predictions, and the combined statistical and theoretical uncertainties for the \bhs prediction. 
   }
  \label{fig:jet_pt_r23jets}
\end{figure*}

\begin{figure*} \centering
  \includegraphics[width=.38\textwidth]{\figurespath/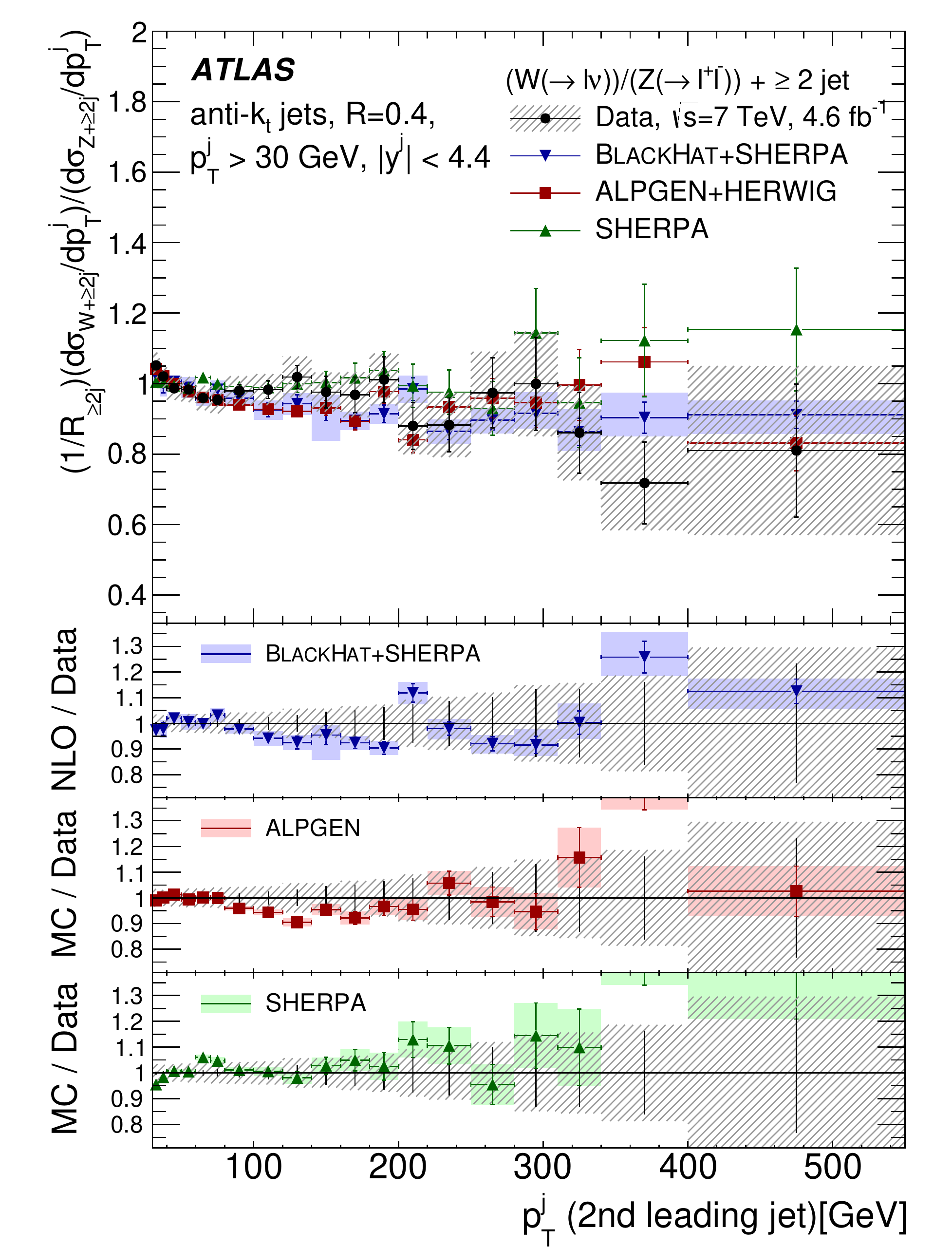}
  \includegraphics[width=.38\textwidth]{\figurespath/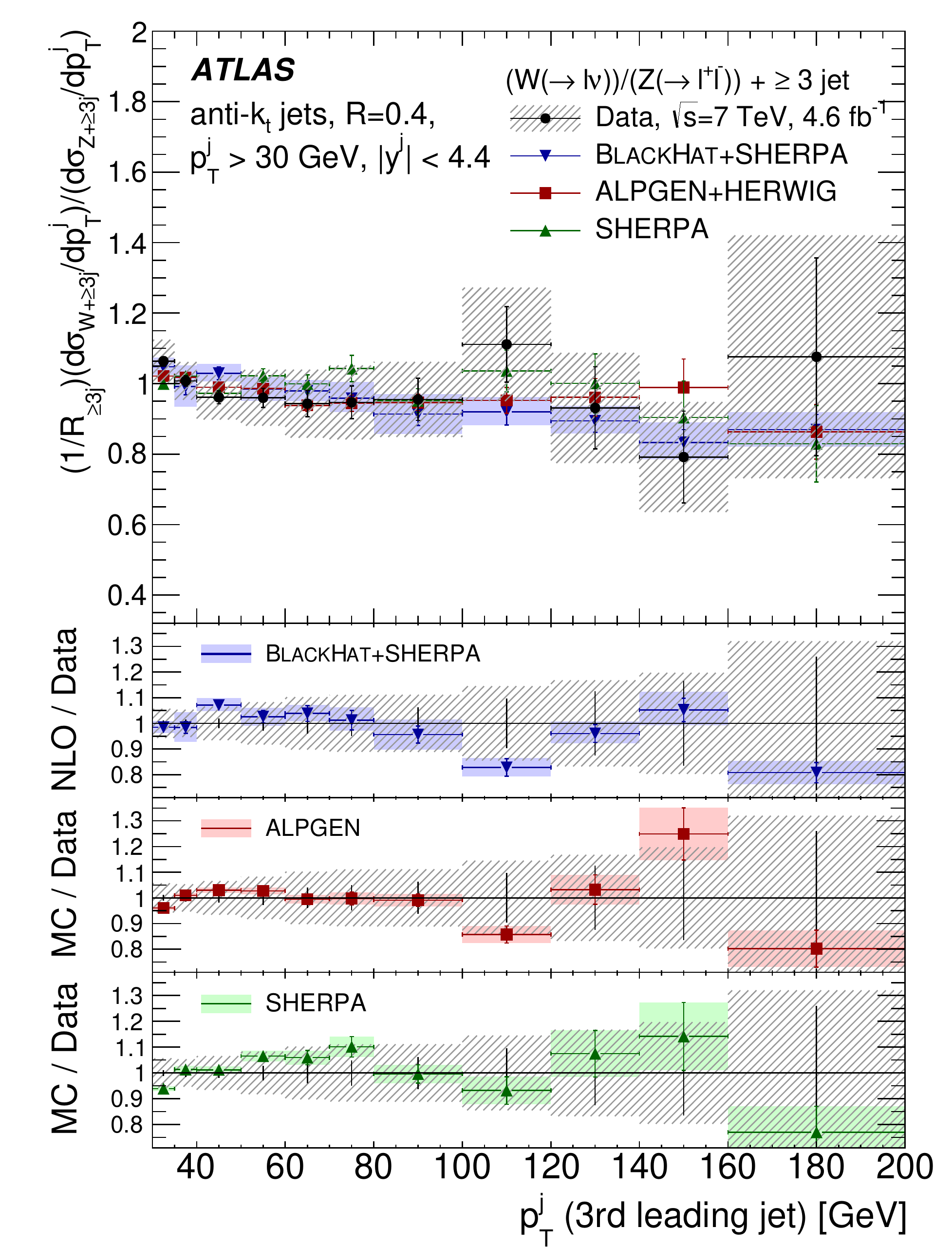}
  \caption{The ratio of \Wjets and \Zjets production cross sections, \rjets, 
  normalized as described in the text versus the second-leading-jet transverse momentum, \ptj, for \njets$\ge$ 2
    (left) and versus the third-leading-jet \pt for \njets $\ge 3$ (right). The
    electron and muon channel measurements are combined as described
    in the text.  Ratios of the \bhs NLO
    calculation and the {\alp} and {\she} generators  to the data are shown in the
    lower panels.
    Vertical error bars show the respective statistical uncertainties. 
    The hatched error band shows statistical and systematic uncertainties added in quadrature for the data.
    The solid error bands show the statistical uncertainties for the {\alp} and {\she} predictions, and the combined statistical and theoretical uncertainties for the \bhs prediction. 
    }
  \label{fig:jet_npt_r23jets}
\end{figure*}

Figure~\ref{fig:jet_npt_r23jets} shows \rjets versus the
second- and third-leading-jet \pt for \njets$\ge 2$ and \njets$\ge 3$ respectively.
The various predictions agree with the data distributions, 
 given the uncertainties,
 except for small deviations in the second-leading-jet \pt for $\njets \geq 2$.

The next kinematic observable studied is \stj, the scalar sum of all
jet transverse momenta in the event.  
This observable is often used in searches for new high-mass particles.
Figure~\ref{fig:inclusive_r2jets} shows \rjets versus \stj 
for $\njets = 2$ and $\njets \ge$ 2, while Fig.~\ref{fig:inclusive_r3jets}
shows \rjets versus \stj for $\njets = 3$ and $\njets \ge$ 3.  
At the lowest values of \stj the predicted distributions are different from 
the measured distributions,
particularly for \she, but in the higher-\stj region
the theoretical predictions describe the data well.
The central value of the fixed-order \bhs calculation does not reproduce the
\stj distributions for \Wjets and \Zjets separately as well as the inclusive calculation,
corroborating the previous observations in Refs.~\cite{Aad:2013ysa,ATLASWjets2014}. 
The tensions 
are due to the missing higher-order 
contributions which cancel almost completely in \rjets.

\begin{figure*} \centering
  \includegraphics[width=.38\textwidth]{\figurespath/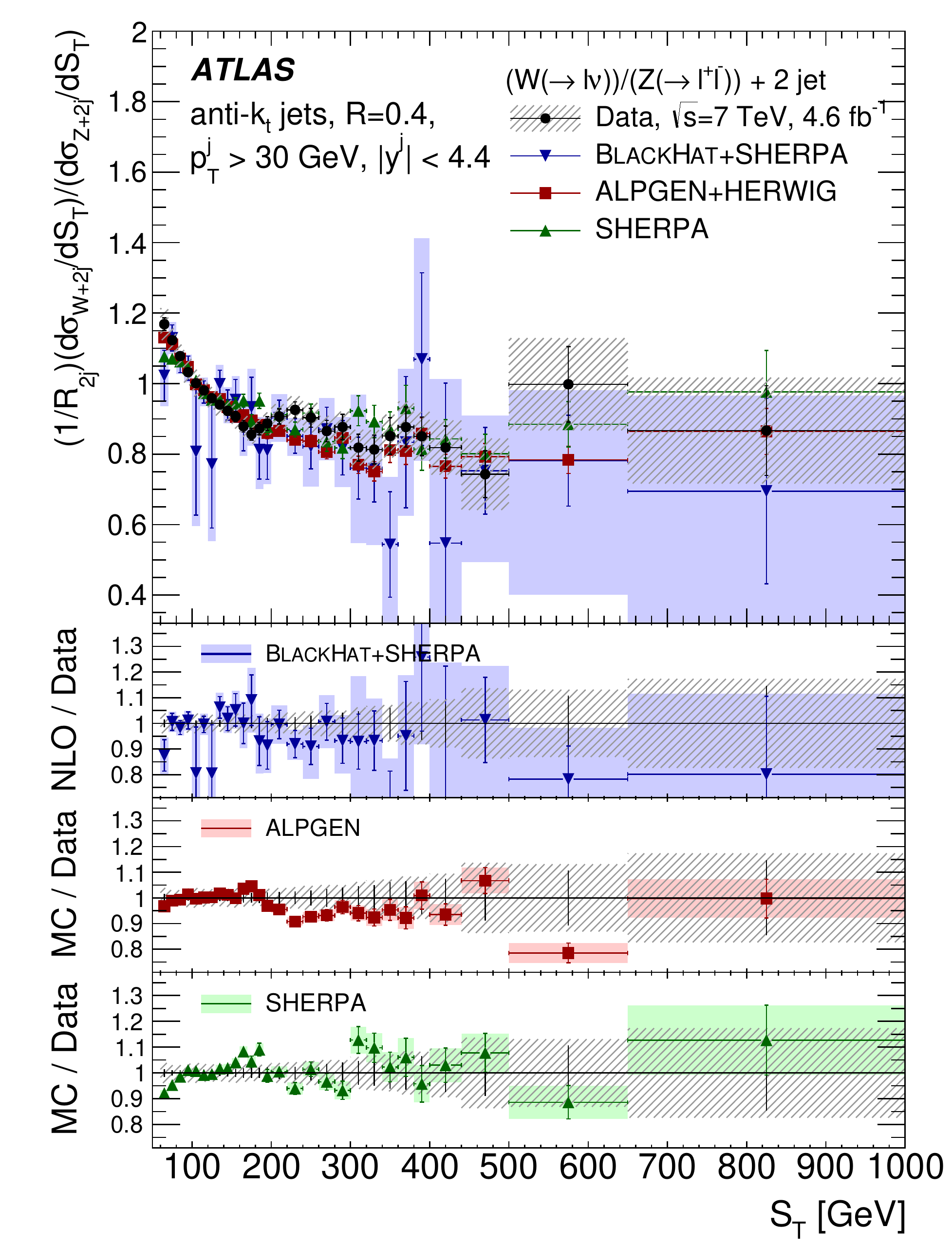}
  \includegraphics[width=.38\textwidth]{\figurespath/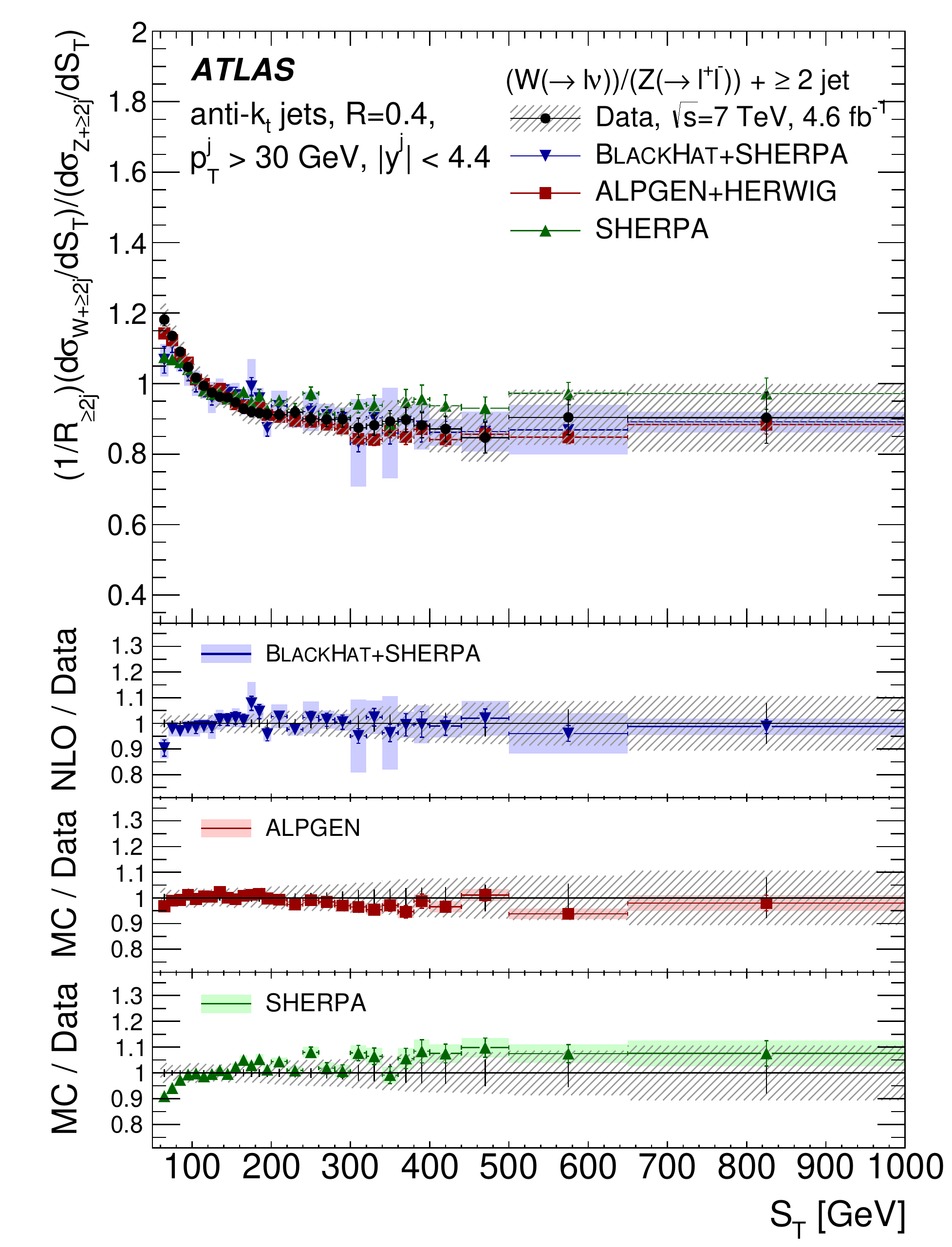}
  \caption{The ratio of \Wjets and \Zjets production cross sections, \rjets,
   normalized as described in the text versus the scalar sum \pt of jets, \stj, for \njets = 2 (left) and $\ge$ 2
    (right). The electron and muon channel measurements are combined as
    described in the text.  Ratios of the
    \bhs NLO calculation and the {\alp} and {\she} generators  to the data are
    shown in the lower panels.
    Vertical error bars show the respective statistical uncertainties. 
    The hatched error band shows statistical and systematic uncertainties added in quadrature for the data.
    The solid error bands show the statistical uncertainties for the {\alp} and {\she} predictions, and the combined statistical and theoretical uncertainties for the \bhs prediction. 
    }
  \label{fig:inclusive_r2jets}
\end{figure*}

\begin{figure*} \centering
  \includegraphics[width=.38\textwidth]{\figurespath/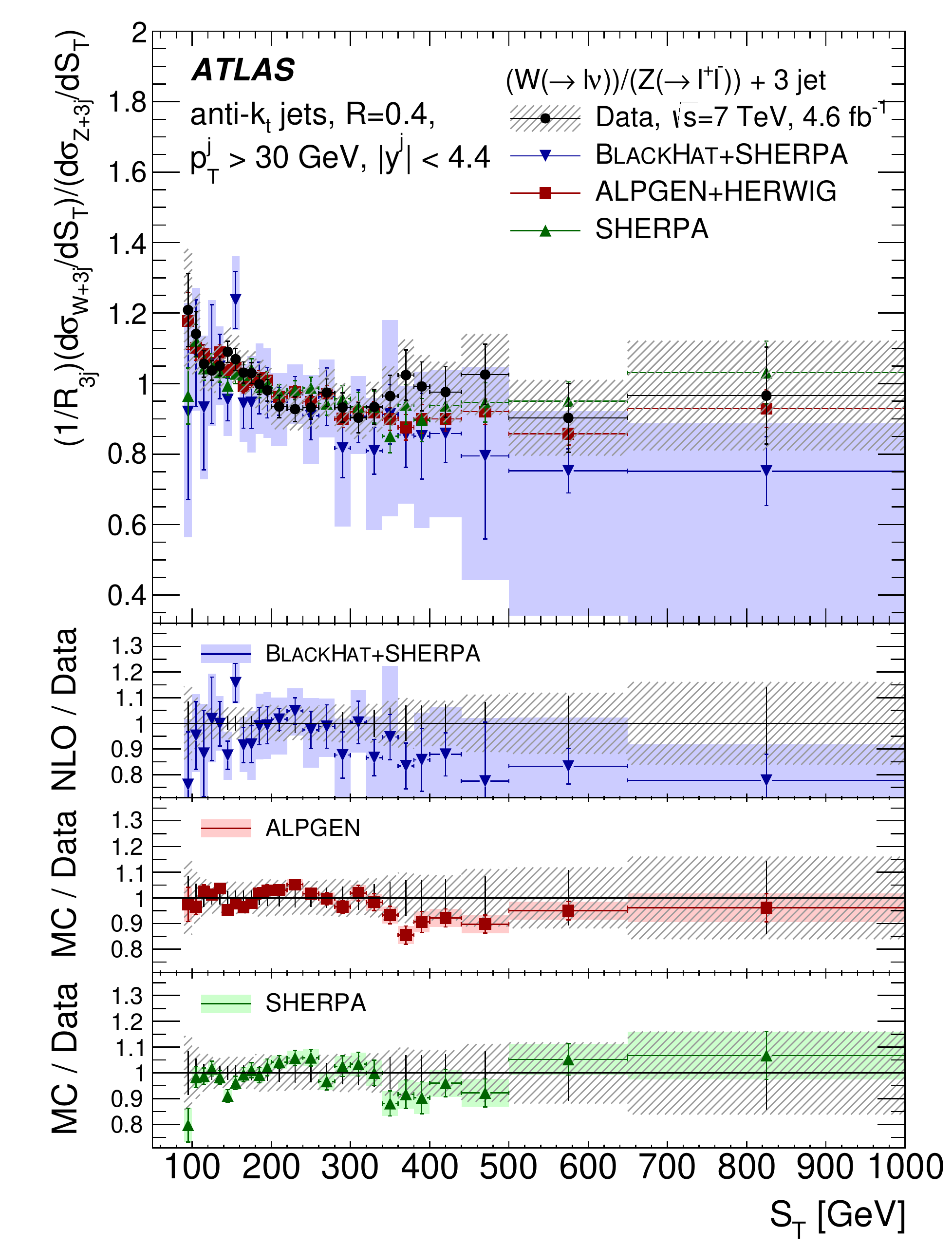}
  \includegraphics[width=.38\textwidth]{\figurespath/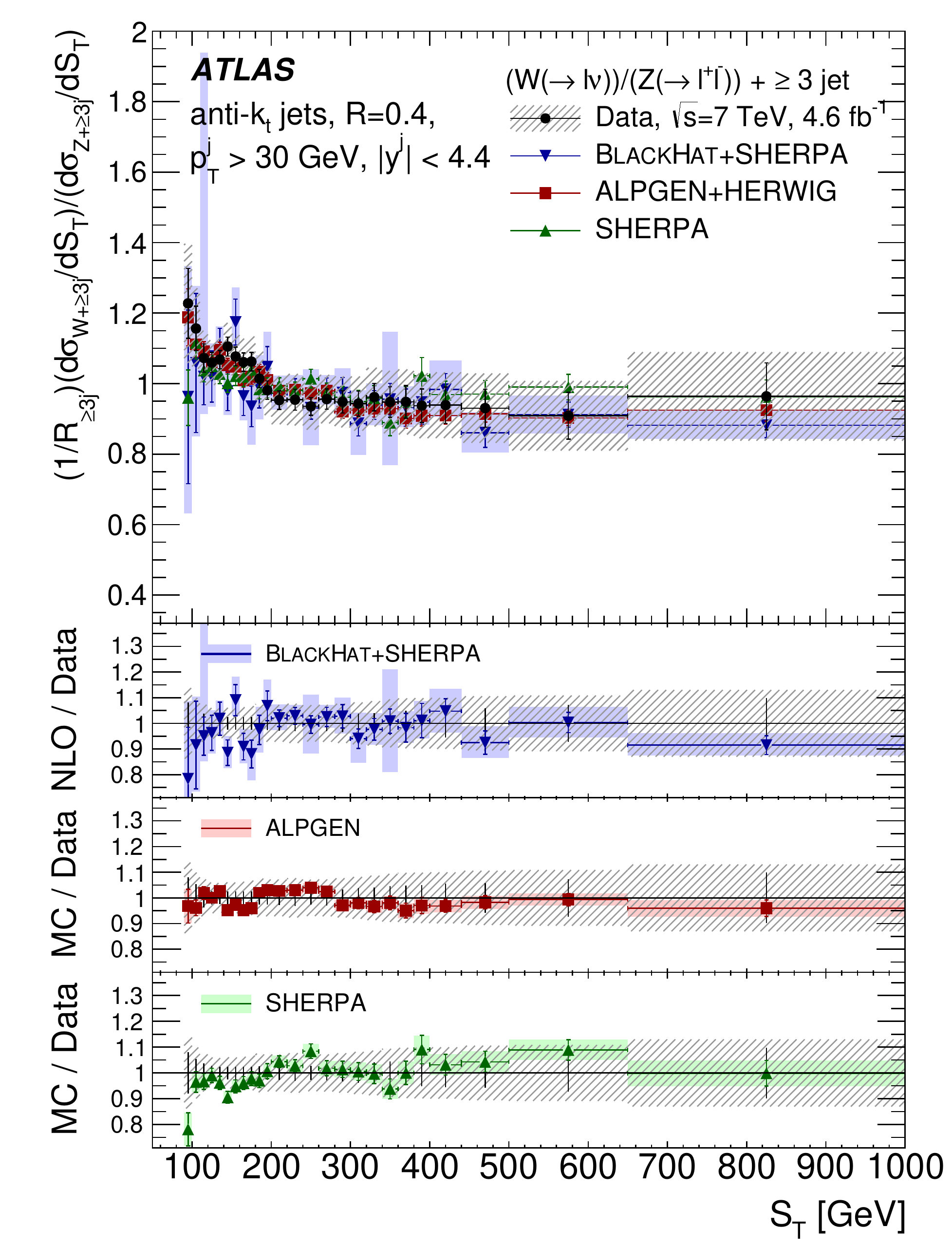}
  \caption{\rjets normalized as described in the text versus the scalar sum \pt of jets, \stj for \njets = 3 (left) and $\ge$ 3
    (right). The electron and muon channel measurements are combined
    as described in the text.  Ratios of the
    \bhs NLO calculation and the {\alp} and {\she} generators  to the data are
    shown in the lower panels.
    Vertical error bars show the respective statistical uncertainties. 
    The hatched error band shows statistical and systematic uncertainties added in quadrature for the data.
    The solid error bands show the statistical uncertainties for the {\alp} and {\she} predictions, and the combined statistical and theoretical uncertainties for the \bhs prediction. 
    }
  \label{fig:inclusive_r3jets}
\end{figure*}

Figure~\ref{fig:dijet_rphi} shows the 
separation
$\Delta R_{\rm j1,j2}$ 
and the azimuthal angular distance $\Delta\phi_{\rm j1,j2}$ between 
the two leading jets, and Fig.~\ref{fig:dijet_minv} shows their
invariant mass $m_{\rm 12}$ for \njets $\ge$ 2.  
At the lowest $\Delta R_{\rm j1,j2}$ and $m_{\rm 12}$ values, the predicted shapes differ from the measured ones. 
This is interpreted as a weak sensitivity to non-perturbative effects
enhancing the difference in soft QCD radiation between $W$ and $Z$ events, but
not cancelling completely in \rjets.  

\begin{figure*} \centering
  \includegraphics[width=.38\textwidth]{\figurespath/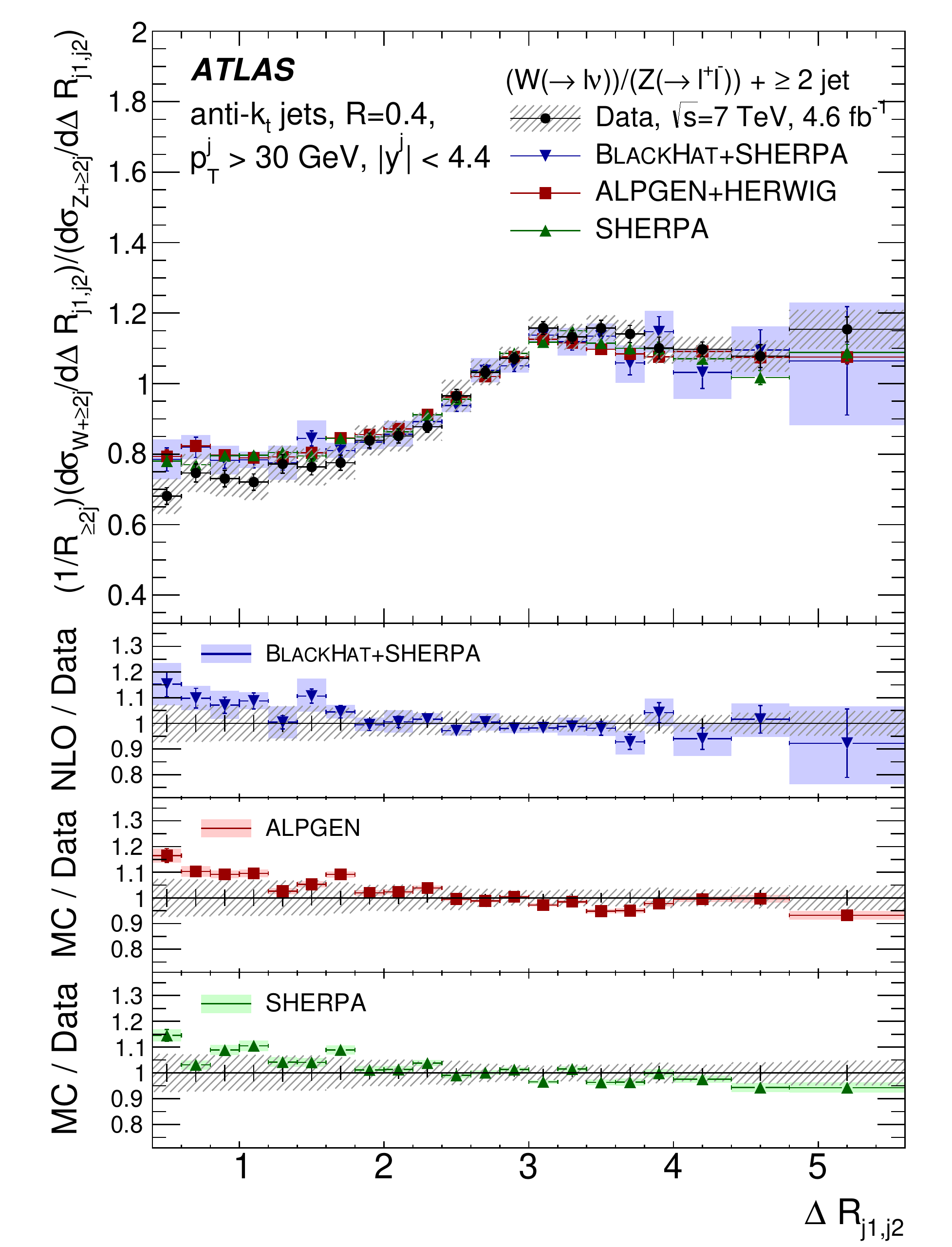}
  \includegraphics[width=.38\textwidth]{\figurespath/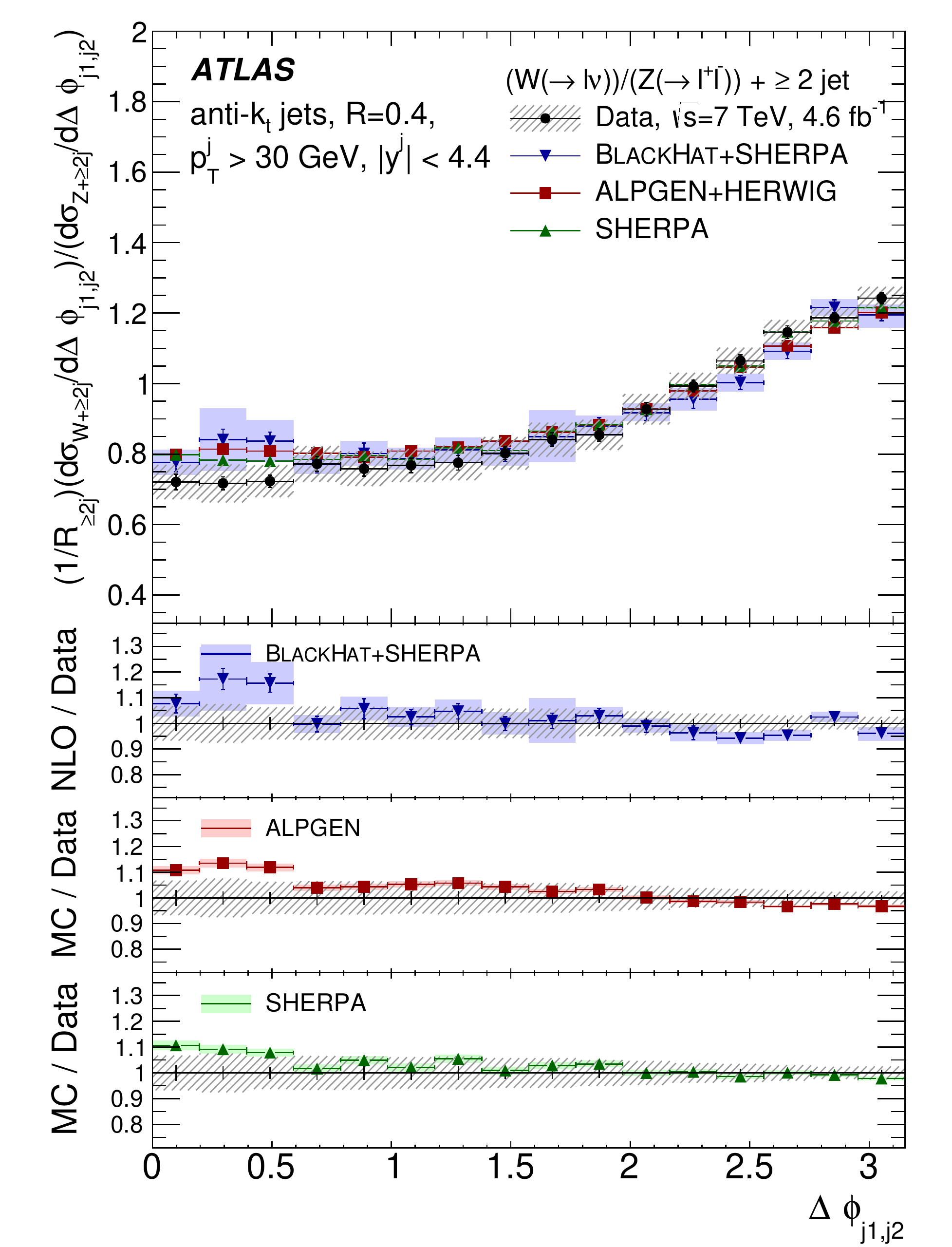}
  \caption{The ratio of \Wjets and \Zjets production cross sections, \rjets, 
  normalized as described in the text versus the dijet angular separation, $\Delta R_{\rm j1,j2}$, (left) and the distance in $\phi$,
    $\Delta\phi_{\rm j1,j2}$, (right) for \njets $\ge 2$. The electron and muon channel
    measurements are combined as described in
    the text. Ratios of the \bhs
    NLO calculation and the {\alp} and {\she} generators to the data are shown in
    the lower panels. 
    Vertical error bars show the respective statistical uncertainties. 
    The hatched error band shows statistical and systematic uncertainties added in quadrature for the data.
    The solid error bands show the statistical uncertainties for the {\alp} and {\she} predictions, and the combined statistical and theoretical uncertainties for the \bhs prediction. 
    }
  \label{fig:dijet_rphi}
\end{figure*}

\begin{figure*} \centering
  \includegraphics[width=.38\textwidth]{\figurespath/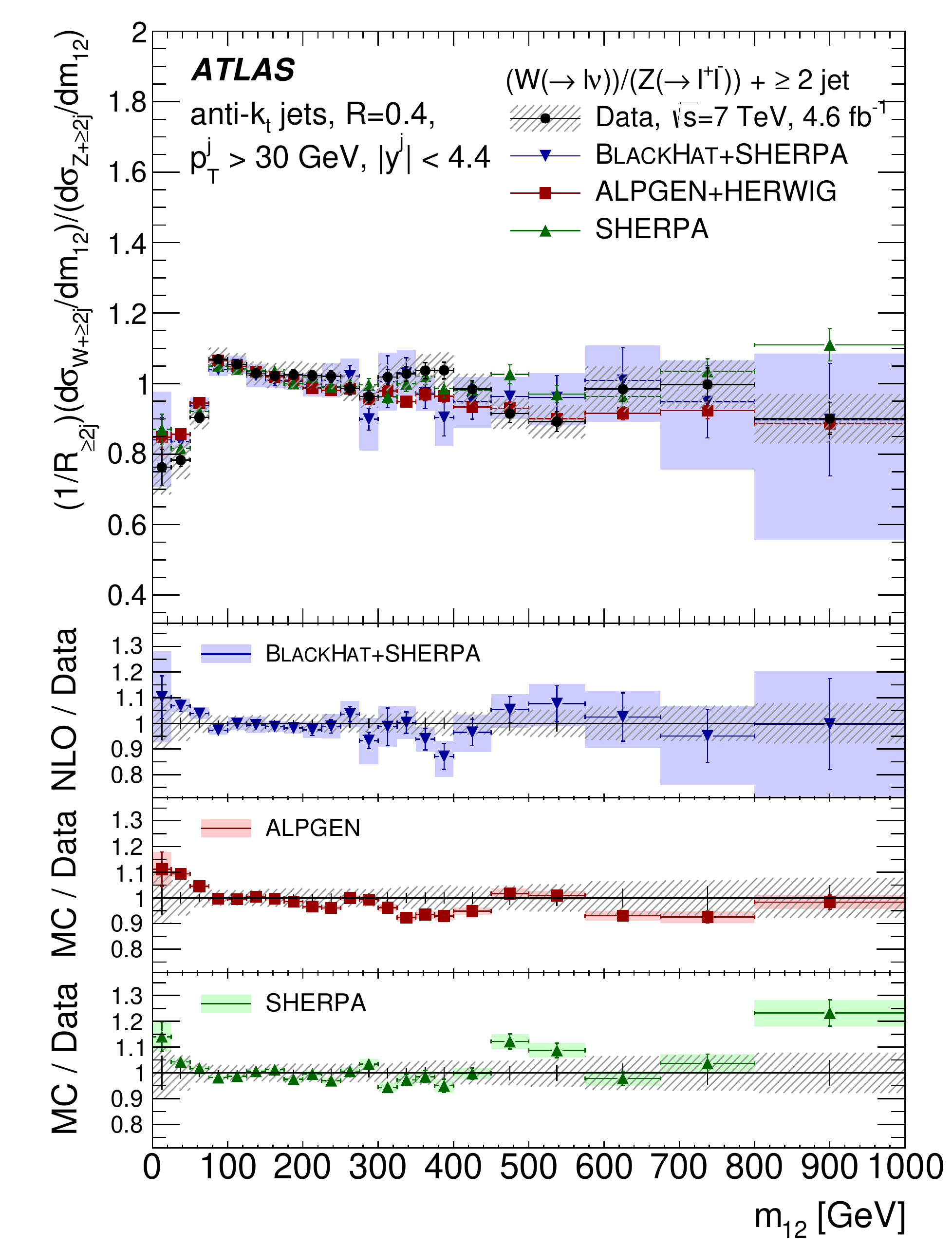}
  \caption{The ratio of \Wjets and \Zjets production cross sections, \rjets, 
  normalized as described in the text 
    versus the dijet invariant mass, $m_{\rm 12}$, for \njets $\ge 2$. The electron and
    muon channel measurements are combined as described in
    the text.  Ratios of the \bhs
    NLO calculation and the {\alp} and {\she} generators  to the data are shown in
    the lower panels.
    Vertical error bars show the respective statistical uncertainties. 
    The hatched error band shows statistical and systematic uncertainties added in quadrature for the data.
    The solid error bands show the statistical uncertainties for the {\alp} and {\she} predictions, and the combined statistical and theoretical uncertainties for the \bhs prediction. 
    }
  \label{fig:dijet_minv}
\end{figure*}

Figure~\ref{fig:jet_rap_r12jets} shows the leading-jet rapidity
for \njets$\ge 1$, and the second-leading-jet rapidity for \njets
$\ge 2$, while Fig.~\ref{fig:jet_rap_r3jets} shows the third-leading-jet 
rapidity for \njets $\ge 3$.  
The different trends between predictions at high leading-jet 
rapidity can be due to the effects of the parton shower 
and, in some cases, different PDF sets.
These effects, which do not cancel completely in \rjets,
are moderated by the presence of extra jets.

\begin{figure*} \centering
  \includegraphics[width=.38\textwidth]{\figurespath/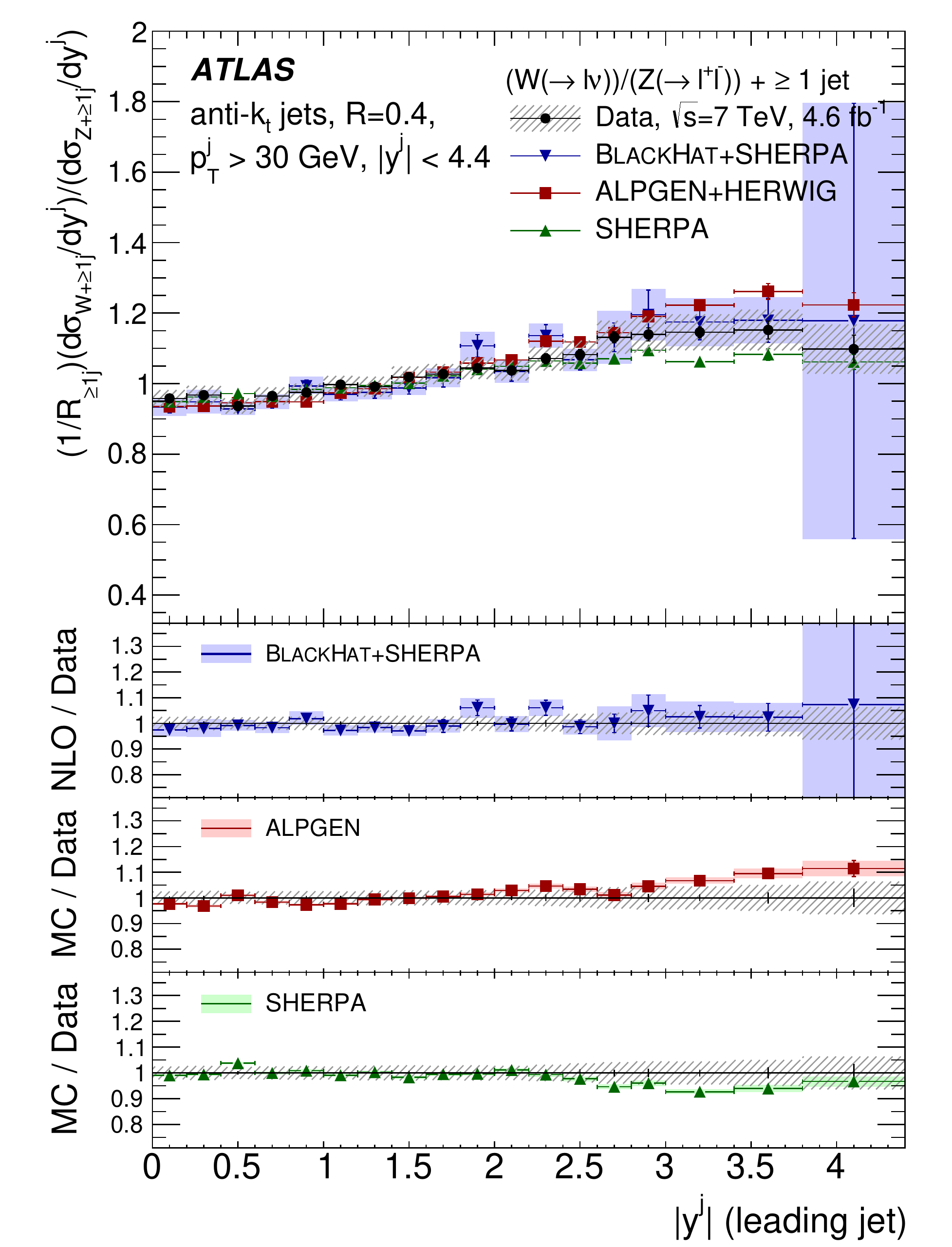}
  \includegraphics[width=.38\textwidth]{\figurespath/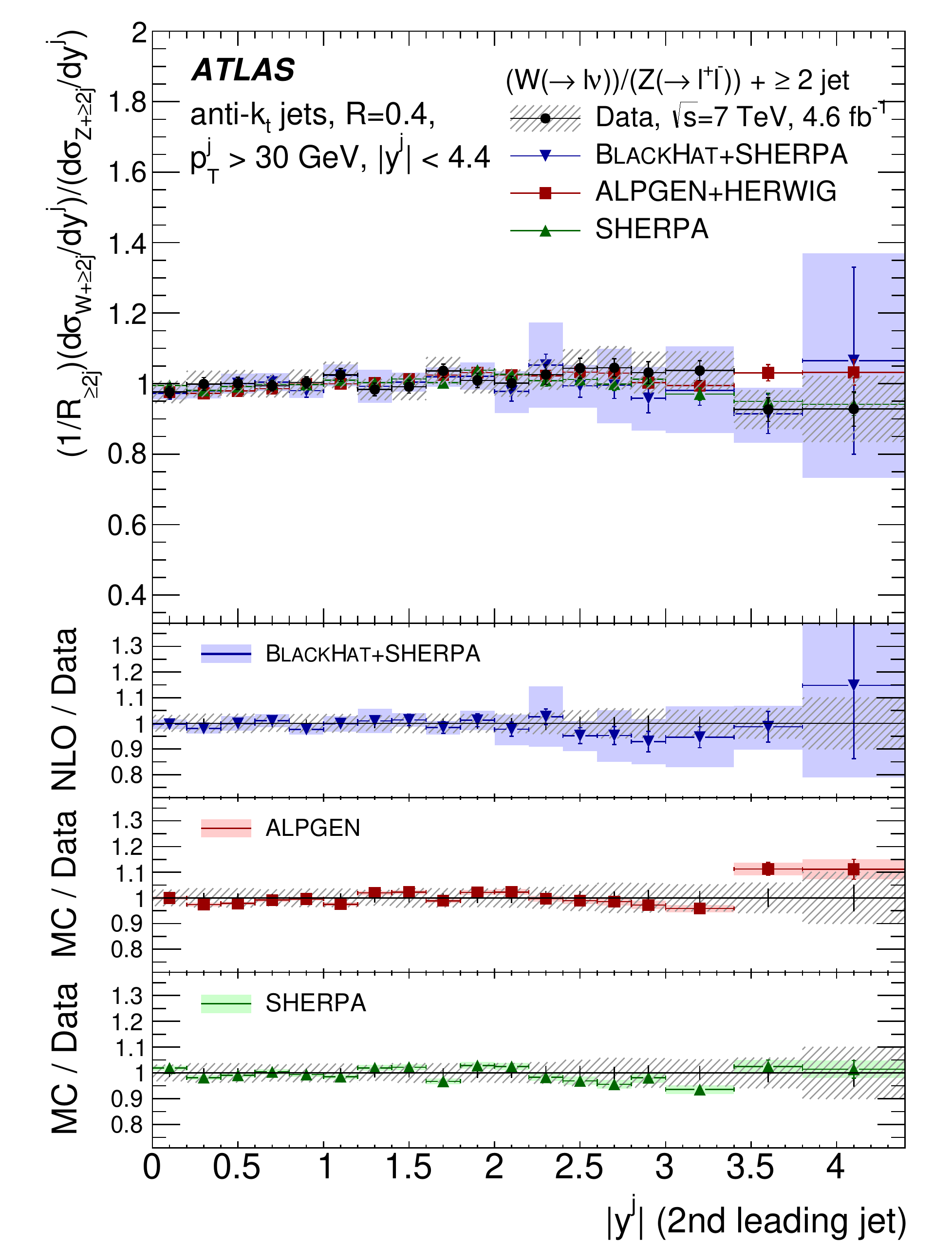}
  \caption{The ratio of \Wjets and \Zjets production cross sections, \rjets, 
  normalized as described in the text versus the leading-jet rapidity, $y^j$, for \njets $\ge
    1$ (left) and second-leading-jet $y$  for \njets $\ge 2$ (right). The
    electron and muon channel measurements are combined as described
    in the text.  Ratios of the \bhs NLO
    calculation and the {\alp} and {\she} generators  to the data are shown in the
    lower panels.
    Vertical error bars show the respective statistical uncertainties. 
    The hatched error band shows statistical and systematic uncertainties added in quadrature for the data.
    The solid error bands show the statistical uncertainties for the {\alp} and {\she} predictions, and the combined statistical and theoretical uncertainties for the \bhs prediction. 
    }
  \label{fig:jet_rap_r12jets}
\end{figure*}

\begin{figure*} \centering
  \includegraphics[width=.38\textwidth]{\figurespath/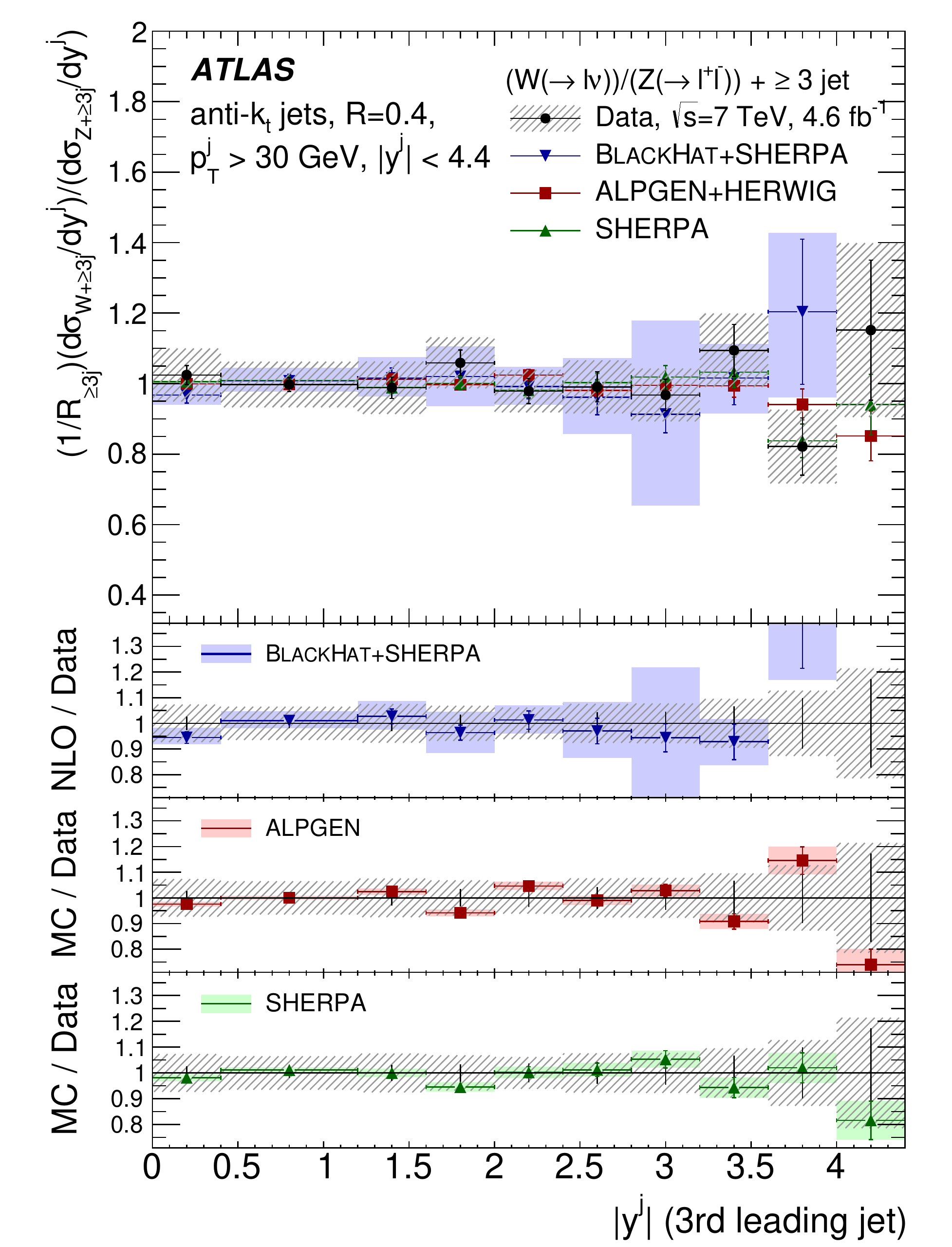}
  \caption{The ratio of \Wjets and \Zjets production cross sections, \rjets, 
  normalized as described in the text versus the third-leading-jet rapidity, $y^j$, for \njets $\ge$
    3. The electron and muon channel measurements are combined as
    described in the text.  Ratios of the
    \bhs NLO calculation and the {\alp} and {\she} generators  to the data are
    shown in the lower panels.
    Vertical error bars show the respective statistical uncertainties. 
    The hatched error band shows statistical and systematic uncertainties added in quadrature for the data.
    The solid error bands show the statistical uncertainties for the {\alp} and {\she} predictions, and the combined statistical and theoretical uncertainties for the \bhs prediction. 
    }
  \label{fig:jet_rap_r3jets}
\end{figure*}

\clearpage
\section{Conclusions\label{Conclusions}}
Measurements of the ratio of \Wjets to \Zjets production cross sections
have been performed by the ATLAS experiment
using a 
data sample of proton--proton collisions corresponding to an integrated luminosity of
$4.6\,\mathrm{fb}^{-1}$ collected
at a centre-of-mass energy of $\sqrt{s}=7\,\tev$ at the LHC.
The data were unfolded to particle level
and compared to predictions from Monte Carlo simulations. 
By being sensitive to differences between \Wjets and \Zjets events,
and through large cancellations of experimental systematic
uncertainties and non-perturbative QCD effects, the \rjets
measurements provide information complementary to individual \Wjets
and \Zjets measurements.
This \rjets measurement significantly improves on previous results 
by probing kinematic distributions for the first time in 
events with jet multiplicity up to four jets.  
It also allows a detailed comparison with
state-of-the-art NLO pQCD Monte Carlo calculations, which
agree well with the observed data except in a few specific regions.
In particular, the \bhs predictions for \rjets at high jet multiplicity and large leading-jet momenta
are validated with this large dataset and are consistent with the results from tuned event generators.
This new measurement highlights the success of recent theoretical advances
and the opportunity for further tuning to 
improve the description of the
production of vector bosons in association with jets.


\section*{Acknowledgments}

We thank CERN for the very successful operation of the LHC, as well as the
support staff from our institutions without whom ATLAS could not be
operated efficiently.

We acknowledge the support of ANPCyT, Argentina; YerPhI, Armenia; ARC,
Australia; BMWFW and FWF, Austria; ANAS, Azerbaijan; SSTC, Belarus; CNPq and FAPESP,
Brazil; NSERC, NRC and CFI, Canada; CERN; CONICYT, Chile; CAS, MOST and NSFC,
China; COLCIENCIAS, Colombia; MSMT CR, MPO CR and VSC CR, Czech Republic;
DNRF, DNSRC and Lundbeck Foundation, Denmark; EPLANET, ERC and NSRF, European Union;
IN2P3-CNRS, CEA-DSM/IRFU, France; GNSF, Georgia; BMBF, DFG, HGF, MPG and AvH
Foundation, Germany; GSRT and NSRF, Greece; ISF, MINERVA, GIF, I-CORE and Benoziyo Center,
Israel; INFN, Italy; MEXT and JSPS, Japan; CNRST, Morocco; FOM and NWO,
Netherlands; BRF and RCN, Norway; MNiSW and NCN, Poland; GRICES and FCT, Portugal; MNE/IFA, Romania; MES of Russia and ROSATOM, Russian Federation; JINR; MSTD,
Serbia; MSSR, Slovakia; ARRS and MIZ\v{S}, Slovenia; DST/NRF, South Africa;
MINECO, Spain; SRC and Wallenberg Foundation, Sweden; SER, SNSF and Cantons of
Bern and Geneva, Switzerland; NSC, Taiwan; TAEK, Turkey; STFC, the Royal
Society and Leverhulme Trust, United Kingdom; DOE and NSF, United States of
America.

The crucial computing support from all WLCG partners is acknowledged
gratefully, in particular from CERN and the ATLAS Tier-1 facilities at
TRIUMF (Canada), NDGF (Denmark, Norway, Sweden), CC-IN2P3 (France),
KIT/GridKA (Germany), INFN-CNAF (Italy), NL-T1 (Netherlands), PIC (Spain),
ASGC (Taiwan), RAL (UK) and BNL (USA) and in the Tier-2 facilities
worldwide.

\providecommand{\href}[2]{#2}\begingroup\raggedright\endgroup


\onecolumn
\clearpage
\begin{flushleft}
{\Large The ATLAS Collaboration}

\bigskip

G.~Aad$^{\rm 84}$,
B.~Abbott$^{\rm 112}$,
J.~Abdallah$^{\rm 152}$,
S.~Abdel~Khalek$^{\rm 116}$,
O.~Abdinov$^{\rm 11}$,
R.~Aben$^{\rm 106}$,
B.~Abi$^{\rm 113}$,
M.~Abolins$^{\rm 89}$,
O.S.~AbouZeid$^{\rm 159}$,
H.~Abramowicz$^{\rm 154}$
% ...
\bigskip
\\
% ...
$^{11}$ Institute of Physics, Azerbaijan Academy of Sciences, Baku, Azerbaijan\\
% ...
$^{84}$ CPPM, Aix-Marseille Universit{\'e} and CNRS/IN2P3, Marseille, France\\
% ...
$^{89}$ Department of Physics and Astronomy, Michigan State University, East Lansing MI, United States of America\\
% ...
$^{106}$ Nikhef National Institute for Subatomic Physics and University of Amsterdam, Amsterdam, Netherlands\\
% ...
$^{112}$ Homer L. Dodge Department of Physics and Astronomy, University of Oklahoma, Norman OK, United States of America\\
$^{113}$ Department of Physics, Oklahoma State University, Stillwater OK, United States of America\\
% ...
$^{116}$ LAL, Universit{\'e} Paris-Sud and CNRS/IN2P3, Orsay, France\\
% ...
$^{152}$ Institute of Physics, Academia Sinica, Taipei, Taiwan\\
% ...
$^{154}$ Raymond and Beverly Sackler School of Physics and Astronomy, Tel Aviv University, Tel Aviv, Israel\\
% ...
$^{159}$ Department of Physics, University of Toronto, Toronto ON, Canada\\
% ...
\end{flushleft}



\begin{thebibliography}{10}

\bibitem{Abouzaid:2003is}
E.\,Abouzaid and H.~J.\,Frisch,
  \href{http://dx.doi.org/10.1103/PhysRevD.68.033014}{Phys.\,Rev.\,D {\bfseries
  68} (2003) 033014},
\href{http://arxiv.org/abs/hep-ph/0303088}{{\ttfamily arXiv:hep-ph/0303088
  [hep-ph]}}.

\bibitem{Aad:2011xn}
{ATLAS Collaboration},
  \href{http://dx.doi.org/10.1016/j.physletb.2012.01.042}{Phys.\,Lett\,B.
  {\bfseries 708} (2012) 221--240},
\href{http://arxiv.org/abs/1108.4908}{{\ttfamily arXiv:1108.4908 [hep-ex]}}.

\bibitem{Chatrchyan:2011ne}
{CMS Collaboration}, \href{http://dx.doi.org/10.1007/JHEP01(2012)010}{JHEP
  {\bfseries 1201} (2012) 010},
\href{http://arxiv.org/abs/1110.3226}{{\ttfamily arXiv:1110.3226 [hep-ex]}}.

\bibitem{ATLASWjets2014}
{ATLAS Collaboration}, Eur. Phys. J. C. \href{http://arxiv.org/abs/1409.8639}{arXiv:1409.8639} [hep-ex] (submitted)

\bibitem{Aad:2013ysa}
{ATLAS Collaboration}, \href{http://dx.doi.org/10.1007/JHEP07(2013)032}{JHEP
  {\bfseries 1307} (2013) 032},
\href{http://arxiv.org/abs/1304.7098}{{\ttfamily arXiv:1304.7098 [hep-ex]}}.

\bibitem{Aad:2008zzm}
{ATLAS Collaboration},
\href{http://dx.doi.org/10.1088/1748-0221/3/08/S08003}{JINST {\bfseries 3}
  (2008) S08003}.

\bibitem{Mangano:2002ea}
M.~L.\,Mangano {et~al.},
  \href{http://dx.doi.org/10.1088/1126-6708/2003/07/001}{JHEP {\bfseries 0307}
  (2003) 001},
\href{http://arxiv.org/abs/hep-ph/0206293}{{\ttfamily arXiv:hep-ph/0206293
  [hep-ph]}}.

\bibitem{Corcella:2000bw}
G.\,Corcella {et~al.},
  \href{http://dx.doi.org/10.1088/1126-6708/2001/01/010}{JHEP {\bfseries 0101}
  (2001) 010},
\href{http://arxiv.org/abs/hep-ph/0011363}{{\ttfamily arXiv:hep-ph/0011363
  [hep-ph]}}.

\bibitem{Butterworth:1996zw}
J.\,Butterworth,\,J.~R.\,Forshaw,\, and M.\,Seymour,
  \href{http://dx.doi.org/10.1007/s002880050286}{Z.\,Phys.\,C {\bfseries 72}
  (1996) 637--646},
\href{http://arxiv.org/abs/hep-ph/9601371}{{\ttfamily arXiv:hep-ph/9601371
  [hep-ph]}}.

\bibitem{Golonka:2006tw}
P.\,Golonka and Z.\,Was,
  \href{http://dx.doi.org/10.1140/epjc/s10052-006-0205-3}{Eur.\,Phys.\,J.\,C
  {\bfseries 50} (2007) 53--62},
\href{http://arxiv.org/abs/hep-ph/0604232}{{\ttfamily arXiv:hep-ph/0604232
  [hep-ph]}}.

\bibitem{Pumplin:2002vw}
J.\,Pumplin {et~al.},
  \href{http://dx.doi.org/10.1088/1126-6708/2002/07/012}{JHEP {\bfseries 0207}
  (2002) 012},
\href{http://arxiv.org/abs/hep-ph/0201195}{{\ttfamily arXiv:hep-ph/0201195
  [hep-ph]}}.

\bibitem{ATLAS:2011gmi}
{ATLAS Collaboration},  ATL-PHYS-PUB-2011-008, 2011.
\newblock \url{http://cdsweb.cern.ch/record/1345343}.

\bibitem{Sjostrand:2006za}
T.\,Sj{\"{o}}strand,\,S.\,Mrenna,\, and P.\,Z.\,Skands,
  \href{http://dx.doi.org/10.1088/1126-6708/2006/05/026}{JHEP {\bfseries 0605}
  (2006) 026},
\href{http://arxiv.org/abs/hep-ph/0603175}{{\ttfamily arXiv:hep-ph/0603175
  [hep-ph]}}.

\bibitem{perugia}
{P.\,Z.\,Skands},
  \href{http://dx.doi.org/10.1103/PhysRevD.82.074018}{Phys.\,Rev.\,D {\bfseries
  82} (2010) 074018},
\href{http://arxiv.org/abs/1005.3457}{{\ttfamily arXiv:1005.3457 [hep-ph]}}.

\bibitem{Gleisberg:2008ta}
T.\,Gleisberg {et~al.},
  \href{http://dx.doi.org/10.1088/1126-6708/2009/02/007}{JHEP {\bfseries 0902}
  (2009) 007},
\href{http://arxiv.org/abs/0811.4622}{{\ttfamily arXiv:0811.4622 [hep-ph]}}.

\bibitem{Hoeche:2012yf}
S.\,Hoeche,\,F.\,Krauss,\,M.\,Sch{\"o}nherr,\, and F.\,Siegert,
  \href{http://dx.doi.org/10.1007/JHEP04(2013)027}{JHEP {\bfseries 1304} (2013)
  027},
\href{http://arxiv.org/abs/1207.5030}{{\ttfamily arXiv:1207.5030 [hep-ph]}}.

\bibitem{Lai:2010vv}
H.-L.\,Lai {et~al.},
  \href{http://dx.doi.org/10.1103/PhysRevD.82.074024}{Phys.\,Rev.\,D {\bfseries
  82} (2010) 074024},
\href{http://arxiv.org/abs/1007.2241}{{\ttfamily arXiv:1007.2241 [hep-ph]}}.

\bibitem{Alioli:2010xd}
S.\,Alioli,\,P.\,Nason,\,C.\,Oleari,\, and E.\,Re,
  \href{http://dx.doi.org/10.1007/JHEP06(2010)043}{JHEP {\bfseries 1006} (2010)
  043},
\href{http://arxiv.org/abs/1002.2581}{{\ttfamily arXiv:1002.2581 [hep-ph]}}.

\bibitem{Kersevan:2004yg}
B.~P.\,Kersevan and E.\,Richter-Was,
  \href{http://dx.doi.org/10.1016/j.cpc.2012.10.032}{Comput.\,Phys.\,Commun.
  {\bfseries 184} (2013) 919--985},
\href{http://arxiv.org/abs/hep-ph/0405247}{{\ttfamily arXiv:hep-ph/0405247
  [hep-ph]}}.

\bibitem{Sherstnev:2007nd}
A.\,Sherstnev and R.\,Thorne,
  \href{http://dx.doi.org/10.1140/epjc/s10052-008-0610-x}{Eur.\,Phys.\,J.\,C
  {\bfseries 55} (2008) 553--575},
\href{http://arxiv.org/abs/0711.2473}{{\ttfamily arXiv:0711.2473 [hep-ph]}}.

\bibitem{Agostinelli:2002hh}
{S. Agostinelli, et al. ({\sc Geant4} Collaboration)},
\href{http://dx.doi.org/10.1016/S0168-9002(03)01368-8}{Nucl.\,Instrum.\,Meth.\,A
  {\bfseries 506} (2003) 250--303}.

\bibitem{Aad:2010ah}
{ATLAS Collaboration},
  \href{http://dx.doi.org/10.1140/epjc/s10052-010-1429-9}{Eur.\,Phys.\,J.\,C
  {\bfseries 70} (2010) 823--874},
\href{http://arxiv.org/abs/1005.4568}{{\ttfamily arXiv:1005.4568
  [physics.ins-det]}}.

\bibitem{Melnikov:2006kv}
K.\,Melnikov and F.\,Petriello,
  \href{http://dx.doi.org/10.1103/PhysRevD.74.114017}{Phys.\,Rev.\,D {\bfseries
  74} (2006) 114017},
\href{http://arxiv.org/abs/hep-ph/0609070}{{\ttfamily arXiv:hep-ph/0609070
  [hep-ph]}}.

\bibitem{Martin:2009iq}
A.\,Martin,\,W.\,Stirling,\,R.\,Thorne,\, and G.\,Watt,
  \href{http://dx.doi.org/10.1140/epjc/s10052-009-1072-5}{Eur.\,Phys.\,J.\,C
  {\bfseries 63} (2009) 189--285},
\href{http://arxiv.org/abs/0901.0002}{{\ttfamily arXiv:0901.0002 [hep-ph]}}.

\bibitem{Cacciari:2011hy}
M.\,Cacciari {et~al.},
  \href{http://dx.doi.org/10.1016/j.physletb.2012.03.013}{Phys.\,Lett.\,B
  {\bfseries 710} (2012) 612--622},
\href{http://arxiv.org/abs/1111.5869}{{\ttfamily arXiv:1111.5869 [hep-ph]}}.

\bibitem{Baernreuther:2012ws}
P.\,B{\"a}rnreuther,\,M.\,Czakon,\, and A.\,Mitov,
  \href{http://dx.doi.org/10.1103/PhysRevLett.109.132001}{Phys.\,Rev.\,Lett.
  {\bfseries 109} (2012) 132001},
\href{http://arxiv.org/abs/1204.5201}{{\ttfamily arXiv:1204.5201 [hep-ph]}}.

\bibitem{Czakon:2012zr}
M.\,Czakon and A.\,Mitov, \href{http://dx.doi.org/10.1007/JHEP12(2012)054}{JHEP
  {\bfseries 1212} (2012) 054},
\href{http://arxiv.org/abs/1207.0236}{{\ttfamily arXiv:1207.0236 [hep-ph]}}.

\bibitem{Czakon:2012pz}
M.\,Czakon and A.\,Mitov, \href{http://dx.doi.org/10.1007/JHEP01(2013)080}{JHEP
  {\bfseries 1301} (2013) 080},
\href{http://arxiv.org/abs/1210.6832}{{\ttfamily arXiv:1210.6832 [hep-ph]}}.

\bibitem{Czakon:2013goa}
M.\,Czakon,\,P.\,Fiedler,\, and A.\,Mitov,
  \href{http://dx.doi.org/10.1103/PhysRevLett.110.252004}{Phys.\,Rev.\,Lett.
  {\bfseries 110} (2013) 252004},
\href{http://arxiv.org/abs/1303.6254}{{\ttfamily arXiv:1303.6254 [hep-ph]}}.

\bibitem{Czakon:2011xx}
M.\,Czakon and A.\,Mitov,
\href{http://arxiv.org/abs/1112.5675}{{\ttfamily arXiv:1112.5675 [hep-ph]}}.

\bibitem{Campbell:2011bn}
J.~M.\,Campbell,\,R.~K.\,Ellis,\, and C.\,Williams,
  \href{http://dx.doi.org/10.1007/JHEP07(2011)018}{JHEP {\bfseries 1107} (2011)
  018},
\href{http://arxiv.org/abs/1105.0020}{{\ttfamily arXiv:1105.0020 [hep-ph]}}.

\bibitem{Aad:2013ucp}
{ATLAS Collaboration},
  \href{http://dx.doi.org/10.1140/epjc/s10052-013-2518-3}{Eur.\,Phys.\,J.\,C
  {\bfseries 73} (2013) 2518},
\href{http://arxiv.org/abs/1302.4393}{{\ttfamily arXiv:1302.4393 [hep-ex]}}.

\bibitem{Aad:2014fxa}
{ATLAS Collaboration},
  \href{http://dx.doi.org/10.1140/epjc/s10052-014-2941-0}{Eur.\,Phys.\,J.\,C
  {\bfseries 74} (2014) 2941},
\href{http://arxiv.org/abs/1404.2240}{{\ttfamily arXiv:1404.2240 [hep-ex]}}.

\bibitem{Aad:2014zya}
{ATLAS Collaboration}, \href{http://arxiv.org/abs/1404.4562}{{\ttfamily
  arXiv:1404.4562 [hep-ex]}}.
{submitted to Eur.\,Phys.\,J.\,C}.

\bibitem{ele_iso}
{ATLAS Collaboration},  ATL-PHYS-PUB-2011-006, 2011.
\newblock \url{http://cdsweb.cern.ch/record/1345327}.

\bibitem{Aad:2012re}
{ATLAS Collaboration},
  \href{http://dx.doi.org/10.1140/epjc/s10052-011-1844-6}{Eur.\,Phys.\,J.\,C
  {\bfseries 72} (2012) 1844},
\href{http://arxiv.org/abs/1108.5602}{{\ttfamily arXiv:1108.5602 [hep-ex]}}.

\bibitem{Cacciari:2008gp}
M.\,Cacciari,\,G.~P.\,Salam,\, and G.\,Soyez,
  \href{http://dx.doi.org/10.1088/1126-6708/2008/04/063}{JHEP {\bfseries 0804}
  (2008) 063},
\href{http://arxiv.org/abs/0802.1189}{{\ttfamily arXiv:0802.1189 [hep-ph]}}.

\bibitem{Aad:2012ag}
{ATLAS Collaboration},
  \href{http://dx.doi.org/10.1140/epjc/s10052-013-2306-0}{Eur.\,Phys.\,J.\,C
  {\bfseries 73} (2013) 2306},
\href{http://arxiv.org/abs/1210.6210}{{\ttfamily arXiv:1210.6210 [hep-ex]}}.

\bibitem{Aad:2013tea}
{ATLAS Collaboration}, \href{http://dx.doi.org/10.1007/JHEP05(2014)059}{JHEP
  {\bfseries 1405} (2014) 059},
\href{http://arxiv.org/abs/1312.3524}{{\ttfamily arXiv:1312.3524 [hep-ex]}}.

\bibitem{btag}
{ATLAS Collaboration},  {ATLAS-CONF-2011-102}, 2011.
\newblock \url{http://cdsweb.cern.ch/record/1369219}.

\bibitem{Aad:2012qf}
{ATLAS Collaboration},
  \href{http://dx.doi.org/10.1016/j.physletb.2012.03.083}{Phys.\,Lett.\,B
  {\bfseries 711} (2012) 244--263},
\href{http://arxiv.org/abs/1201.1889}{{\ttfamily arXiv:1201.1889 [hep-ex]}}.

\bibitem{D'Agostini:1994zf}
G.\,D'Agostini,
\href{http://dx.doi.org/10.1016/0168-9002(95)00274-X}{Nucl.\,Instrum.\,Meth.\,A
  {\bfseries 362} (1995) 487--498}.

\bibitem{ATLAS:Met}
{ATLAS Collaboration},  ATLAS-CONF-2012-101, 2012.
\newblock \url{http://cdsweb.cern.ch/record/1463915}.

\bibitem{Alwall:2007fs}
J.\,Alwall {et~al.},
  \href{http://dx.doi.org/10.1140/epjc/s10052-007-0490-5}{Eur.\,Phys.\,J.\,C
  {\bfseries 53} (2008) 473--500},
\href{http://arxiv.org/abs/0706.2569}{{\ttfamily arXiv:0706.2569 [hep-ph]}}.

\bibitem{Glazov:2005rn}
A.\,Glazov,
\href{http://dx.doi.org/10.1063/1.2122026}{AIP\,Conf.\,Proc. {\bfseries 792}
  (2005) 237--240}.

\bibitem{Aaron:2009bp}
{H1} Collaboration, F.\,Aaron {et~al.},
  \href{http://dx.doi.org/10.1140/epjc/s10052-009-1128-6}{Eur.\,Phys.\,J.\,C
  {\bfseries 63} (2009) 625--678},
\href{http://arxiv.org/abs/0904.0929}{{\ttfamily arXiv:0904.0929 [hep-ex]}}.

\bibitem{Berger:2009ep}
C.\,Berger {et~al.},
  \href{http://dx.doi.org/10.1103/PhysRevD.80.074036}{Phys.\,Rev.\,D {\bfseries
  80} (2009) 074036},
\href{http://arxiv.org/abs/0907.1984}{{\ttfamily arXiv:0907.1984 [hep-ph]}}.

\bibitem{Berger:2010vm}
C.\,Berger {et~al.},
  \href{http://dx.doi.org/10.1103/PhysRevD.82.074002}{Phys.\,Rev.\,D {\bfseries
  82} (2010) 074002},
\href{http://arxiv.org/abs/1004.1659}{{\ttfamily arXiv:1004.1659 [hep-ph]}}.

\bibitem{Berger:2010zx}
C.\,Berger {et~al.},
  \href{http://dx.doi.org/10.1103/PhysRevLett.106.092001}{Phys.\,Rev.\,Lett.
  {\bfseries 106} (2011) 092001},
\href{http://arxiv.org/abs/1009.2338}{{\ttfamily arXiv:1009.2338 [hep-ph]}}.

\bibitem{Yennie:1961ad}
D.\,Yennie,\,S.~C.\,Frautschi,\, and H.\,Suura,
\href{http://dx.doi.org/10.1016/0003-4916(61)90151-8}{Annals\,Phys. {\bfseries
  13} (1961) 379--452}.

\end{thebibliography}
\end{document}